\documentclass[twocolumn,prd,superscriptaddress,preprintnumbers,nofootinbib]{revtex4}[11pt]
\pdfoutput=1
\usepackage{amsmath,amssymb,graphicx}
\graphicspath{{figs/}}
\usepackage{epsf,verbatim}
\usepackage{hyperref}
\usepackage{comment}
\usepackage{color}
\usepackage{slashed}
\usepackage{subfigure}
\usepackage[usenames,dvipsnames]{xcolor}
\usepackage{comment}
\usepackage{mdwlist, paralist}
\usepackage{rotating}
\usepackage{multirow}

\usepackage[utf8]{inputenc}

\def\to{\rightarrow}

\def\mO{\mathcal{O}}
\def\TeV{~{\mbox{TeV}}}

\def\GeV{~{\mbox{GeV}}}

\def\mL{\mathcal{L}}

\def\tbf{\textbf}

\emergencystretch=\maxdimen
\begin{document}

\title{The Techni-Pati-Salam Composite Higgs}

\author{Giacomo Cacciapaglia}
\thanks{g.cacciapaglia@ipnl.in2p3.fr}
\author{Shahram Vatani}
\thanks{vatani@ipnl.in2p3.fr}
\affiliation{Institut de Physique des 2 Infinis (IP2I), CNRS/IN2P3, UMR5822, F-69622 Villeurbanne, France}
\affiliation{Universit\'e de Lyon, France; Universit\'e Claude Bernard Lyon 1, Lyon, France}

\author{Chen Zhang}
\thanks{czhang@cts.nthu.edu.tw}
\affiliation{Physics Division, National Center for Theoretical Sciences, Hsinchu, Taiwan 300}

\begin{abstract}
Composite Higgs models can be extended to the Planck scale by means of the \emph{partially unified partial compositeness} (PUPC)
framework. We present in detail the Techni-Pati-Salam model, based on a renormalizable gauge theory $SU(8)_{PS}\times SU(2)_L\times SU(2)_R$.
We demonstrate that masses and mixings for all generations of standard model fermions can be obtained via partial compositeness
at low energy, with four-fermion operators mediated by either heavy gauge bosons or scalars.
The strong dynamics is predicted to be that of a confining $Sp(4)_{\rm HC}$ gauge group, with hyper-fermions in the fundamental
and two-index anti-symmetric representations, with fixed multiplicities. This motivates for Lattice studies of the Infra-Red
near-conformal walking phase, with results that may validate or rule out the model. This is the first complete and realistic
attempt at providing an Ultra-Violet completion for composite Higgs models with top partial compositeness.
In the baryon-number conserving vacuum, the theory also predicts a Dark Matter candidate, with mass in the few TeV
range, protected by semi-integer baryon number.
\end{abstract}

\maketitle

\setcounter{equation}{0} \setcounter{footnote}{0}

\section{Introduction}
\label{sec:intro}

The Standard Model (SM) of particle physics~\cite{Glashow:1961tr,Weinberg:1967tq,Salam:1968rm} has withstood all the attempts at discovering signs of New Physics,
with most recently the null results from the LHC experiments. The discovery of a Higgs-like boson~\cite{Aad:2012tfa,Chatrchyan:2012xdj} has further
confirmed the validity of the SM. The main experimental confirmation has come from precise measurements in
the electroweak (EW) sector of the theory, with most prominent results coming from LEP~\cite{Barbieri:2004qk}. What we know with
a precision at the level of per-mille, is that there exist three Goldstone bosons, i.e. the longitudinal polarizations of
the $W^\pm$ and $Z$ gauge bosons, that complement the gauge principle in the SM and provide mass to the 
weak gauge bosons~\cite{Englert:1964et,Higgs:1964ia,Higgs:1964pj,Guralnik:1964eu}. While all experimental results seem to point towards a SM-like Higgs boson, our knowledge of
its properties is still far from the precision achieved in the gauge sector: the couplings of the Higgs boson are only
known at best at the level of $10\%$~\cite{Khachatryan:2016vau}, and the precision will not improve greatly at the end of the LHC 
programme~\cite{Cepeda:2019klc}. This experimental status leaves open the question of the true nature of the discovered Higgs boson.

On the model building side, the SM lacks two key ingredients that play a crucial role in our understanding of our Universe:
gravitational interactions and a Dark Matter candidate. This simple observation points towards the existence of a new physical
scale, ultimately the Planck mass from gravity\footnote{The intrinsic scale of Dark Matter is not known, however the only direct 
evidences derive from gravitational effects.}, thus keeping open the long standing hierarchy problem between the EW scale and such
Ultra-Violet (UV) scale. The presence of an elementary scalar field in the Higgs sector is particularly at odds with the observed
hierarchy, as a scalar mass receives quantum corrections proportional to the new physical scale. The discovery of a
Higgs boson with a mass of $125$~GeV can, therefore, be considered a materialization of the so-called ``naturalness''
problem. A time-honored possibility~\cite{Weinberg:1975gm} is to replace the elementary Higgs sector of the SM with a strong confining
dynamics: the EW scale would therefore be generated dynamically, like the QCD scale, and  the EW symmetry breaking (EWSB)
can be ascribed to a spontaneous chiral symmetry breaking. While the first proposals were essentially Higgless~\cite{Dimopoulos:1979es,Eichten:1979ah}, it was
soon realized that extending the global symmetry of the theory allows the entire Higgs doublet to arise as a
pseudo-Nambu-Goldstone boson (pNGB) of the condensing strong sector~\cite{Kaplan:1983fs}. This new approach
kills two birds with one stone: it explains why the Higgs boson is lighter than other composite states (in agreement
with the null results of New Physics searches at the LHC) and  the ten per-cent agreement of the composite
Higgs couplings to SM predictions, at the price of generating a ``little hierarchy''~\cite{Barbieri:2000gf} between the EW scale
$v=246$~GeV and the compositeness scale. The latter is encoded in the pNGB decay constant $f \approx \mathcal{O} (1)$~TeV.

The nemesis of this approach to the EWSB is the generation of fermion masses~\cite{Dimopoulos:1979es,Eichten:1979ah}: as SM fermion
couplings to the strong sector typically arise via higher dimension operators, generating large masses (i.e., the top mass) is
generically at odds with fulfilling constraints from flavor changing neutral currents (FCNCs). Many palliatives have been proposed: 
among the most remarkable are the presence of an Infra-Red (IR) conformal phase~\cite{Holdom:1981rm} and the
mechanism of fermion Partial Compositeness (PC)~\cite{Kaplan:1991dc}.
The former relies on the property that the strong sector enters a ``walking'' phase~\cite{Cohen:1988sq} right above the
condensation scale, where a large anomalous dimension of the composite Higgs operator is generated, allowing to push the flavor scale high enough without
suppressing the SM fermion mass operators. In the latter, Yukawa-like couplings are replaced by linear mixing of the
SM fermion fields to fermionic composite operators, in such a way that the large anomalous dimensions are associated to 
composite baryonic operators instead of the Higgs one.
This scenario has been revived in the early 2000's thanks to the principle of holography~\cite{Contino:2003ve}, which allowed to relate 
a composite pNGB Higgs in a nearly-conformal theory to a gauge boson in a warped five-dimensional theory.
Composite Higgs models thus merged with Gauge-Higgs unification model building in warped space~\cite{Hosotani:2005nz}, leading
to the definition of a minimal model based on the symmetry breaking SO(5)/SO(4)~\cite{Agashe:2004rs,Agashe:2005dk}, where only the Higgs doublet
populates the pNGB sector of the theory. A lot of work has been devoted in the literature on this scenario, and we refer to the recent
reviews~\cite{Contino:2010rs,Bellazzini:2014yua,Panico:2015jxa,Cacciapaglia:2020kgq} and references therein.
Yet, most of the results in the literature rely on effective field theory (EFT) analyses, both for studying the phenomenology and for developing various
model building aspects of the composite Higgs paradigm. In the case of the flavor issue~\cite{Matsedonskyi:2014iha,Cacciapaglia:2015dsa,Panico:2016ull}, for instance, it has been found that
light top partners are allowed as soon as flavor structures for light generations can be generated at a higher scale
separated from the condensation scale by a near-conformal phase.

In this article, we want to face the daring need for an UV completion for composite Higgs models: this step is crucial in order to base 
all we learned from EFT studies on more solid foundations and to truly understand the origin of flavor physics. 
Following the holography principle, one may be tempted to invoke extra dimensional theories as genuine UV completions~\cite{Agashe:2005vg}.
We do not find this route satisfactory. On the one hand, some basic requirements at the foundation of the original holographic conjecture~\cite{Maldacena:1997re},
like the presence of maximal supersymmetry, are not satisfied in the minimal models studied so far. Holography applied 
outside of supersymmetry and string theory, while proven to be phenomenologically useful even in QCD~\cite{Erlich:2005qh,DaRold:2005mxj,Hirn:2005nr},
is less robust and tested that the original one, mainly due to the lack of calculability in the strongly-coupled side. Example of models based
on more solid supersymmetric dualities can be found in Refs~\cite{Caracciolo:2012je,Marzocca:2013fza}, however these theories 
lack a complete theory of flavor. On the other hand, it is not clear at all if extra dimensional theories are fundamental
because of the mass dimension carried by gauge couplings themselves~\cite{Gies:2003ic,Morris:2004mg}.
An attractive and time-honored route is offered by microscopic gauge-fermion theories, similar to QCD for mesons and hadrons,
defined in terms of a renormalizable and fundamental 4-dimensional gauge theory (we refer the reader to the recent review of this approach in~\cite{Cacciapaglia:2020kgq}). Attempts to build microscopic descriptions of theories
of top PC can be found in Refs~\cite{Barnard:2013zea,Ferretti:2013kya,Vecchi:2015fma}: these analyses lead to the interesting conclusion that there
exists a limited number of theories apt at describing the low energy composite spectrum~\cite{Ferretti:2016upr,Belyaev:2016ftv}.
These models, however, are not genuine UV completions: they are only able to characterize the dynamics of the model below the
condensation scale, where resonances associated to the pNGB Higgs and top PC are needed. Crucial ingredients like the
near conformal dynamics, the origin of the PC couplings and the masses for leptons and light quarks are absent. The theories are
in fact defined in such a way that they are outside of the conformal window, i.e. they do condense at low energy. Lattice studies of 
some of these theories are also available~\cite{DeGrand:2015zxa,Bennett:2017kga,
Ayyar:2018zuk,Ayyar:2018glg,Bennett:2019jzz,Hasenfratz:2016gut}. An alternative way to UV complete fermion PC is to introduce (light) scalar fields 
charged under the confining gauge symmetry~\cite{Sannino:2016sfx,Cacciapaglia:2017cdi}: at the price of giving up naturalness, one potentially obtains 
a complete and fundamental theory of flavor~\cite{Sannino:2017utc}. We should also mention the possibility of bosonic Technicolor~\cite{Samuel:1990dq},
where an elementary Higgs doublet is re-introduced~\cite{Galloway:2016fuo,Agugliaro:2016clv}.
Trying to achieve a complete theory based on gauge and fermion fields is a much more daring task: this would be similar
to the quest for extended Technicolor theories~\cite{Dimopoulos:1979es,Eichten:1979ah} that, despite intense efforts~\cite{Hill:2002ap,Appelquist:2003uu,
Appelquist:2003hn,Appelquist:2004ai}, has not produced any fully realistic model so far.
In Ref.~\cite{Cacciapaglia:2019vce} an attempt has been made to achieve top PC in confining chiral gauge theories.
More recently, large $N_f$ asymptotic safety~\cite{Antipin:2017ebo} has been proposed as a route to the Planck scale 
for gauge-fermion top PC models~\cite{Cacciapaglia:2018avr}, yet the four fermion interactions leading to PC need to be generated by 
mediation of heavy scalars.

In the present work, we follow the route opened in Ref.~\cite{Cacciapaglia:2019dsq} within the \emph{partially unified partial compositeness} (PUPC) framework:
the confining gauge symmetry is partially unified with the SM ones, with the gauge symmetry breaking due to high-scale scalars. In this sense, 
this approach lies in between the early extended Technicolor approaches and theories with scalars, while retaining the ambition of achieving
a complete theory of flavor in a natural way, i.e. without large hierarchies between scalar masses and the Planck scale.
The partial compositeness four-fermion (PC4F) interactions are thus mediated by the massive gauge bosons, as well as by the
heavy scalars, with large anomalous dimensions still needed to achieve unsuppressed low energy operators in the condensed phase~\cite{Contino:2010rs}.
While it is hard to produce accurate low energy predictions, mainly due to the presence of strong coupling regimes in the IR walking phase and
at low energies, we want to demonstrate in detail how a complete theory of flavor can be achieved in this framework. While the general
idea is described in Ref.~\cite{Cacciapaglia:2019dsq}, here we focus specifically on the Techni-Pati-Salam (TPS) model based on a partially unified 
gauge symmetry
$$
\mathcal{G}_{\rm TPS} = SU(8)_{\rm PS} \times SU(2)_L \times SU(2)_R\,.
$$
We will show how to construct a minimal model, which also helps predicting the properties of the microscopic theory underlying the
low energy composite dynamics (that can be studied on the lattice), and the dynamics of the walking phase.
Analysing how flavor structures arise can help better understand the low energy properties of composite models: for
instance, we can show that the multi-scale scenario of Ref.~\cite{Panico:2016ull} cannot be achieved in this framework
and only top partners, i.e. light-ish spin-1/2 resonances associated to the third generation, are possible.

The article is organised as follows. In Section~\ref{sec:gen} we present the general features of the PUPC framework,
and the characteristics that lead us to focus on the TPS model and its symmetry breaking pattern. In 
Section~\ref{sec:tps3} we discuss in detail how the masses for the third generation of SM fermions can be generated, starting from a fundamental
gauge-Yukawa theory at high scale. In particular, we will show how the mass hierarchy between top, bottom, tau and neutrino can be achieved.
In Section~\ref{sec:ext} we investigate the
possibility of extending the construction to the first and second families: we identify the necessary and minimal ingredients needed
to generating all masses and non-trivial Cabibbo-Kobayashi-Maskawa (CKM)~\cite{Cabibbo:1963yz,Kobayashi:1973fv} and Pontecorvo-Maki-Nakagawa-Sakata (PMNS)~\cite{Maki:1962mu} mixing matrices.
We also establish how baryon number conservation can be imposed to avoid proton decay, thus leading to the existence
of a potential Dark Matter candidate.
We offer our conclusion and the perspectives in Section~\ref{sec:dnc}.

\section{General considerations}
\label{sec:gen}

\begin{figure*}[ht]
\includegraphics[width=6.0in]{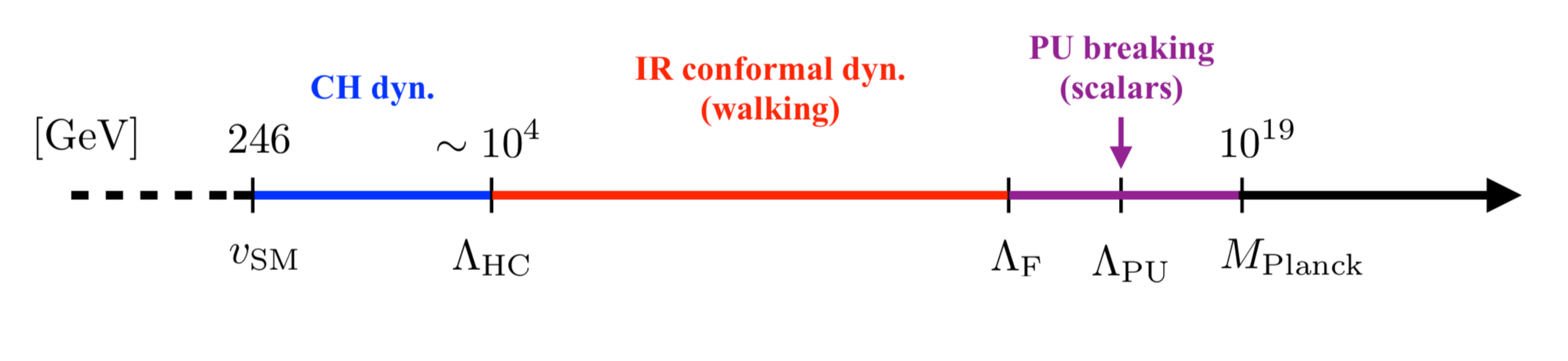}
\caption{\label{fig:1}Schematic representation of the dynamical phases of PUPC models.}
\end{figure*}

The main goal of our PUPC approach~\cite{Cacciapaglia:2019dsq} is to provide a genuine UV completion for composite Higgs models with top partial  compositeness,  which  could  explain  the  origin of  the  partial  compositeness  couplings and flavor physics. The  theory also  needs  to  be  valid  all  the  way  up  to  the  Planck scale,  where  quantum  gravity  effects  become  relevant. To achieve this goal, we require that the theory in the UV consists of a renormalizable gauge-Yukawa theory.
Scalars, therefore, are added with a ``natural'' potential, in the sense that all the dimension-of-mass parameters are not too far from the Planck scale. We remind the reader that this ``naturalness'' principle does not apply to fermion masses. The low energy target is a composite Higgs model with, at least, top partial compositeness.
This implies that the UV theory needs to provide both the couplings to achieve top PC, and an intermediate walking phase to enhance them at low energy: the PUPC model, therefore, needs to pass through several different dynamical phases at various scales, as schematically depicted in Fig.~\ref{fig:1}. Here, we expect the low energy dynamics, above the EW scale, to be that of a confining theory with a typical scale $\Lambda_{\rm HC} \approx 10$~TeV (implying a Higgs pNGB decay constant $f \approx \frac{\Lambda_{\rm HC}}{4 \pi} \approx 1$~TeV). An IR walking phase thus occurs, separating the confinement scale from the scale where flavor physics is generated, $\Lambda_{\rm F}$. How large this scale needs to be depends on the flavor bounds in a specific model, however we expect it to be close to the scale of gauge symmetry breaking of the UV theory. The latter is achieved by giving vacuum expectation values (VEVs) to the scalars in the theory, at a scale $\Lambda_{\rm PU}$, which is allowed to be roughly one loop-factor below $M_{\rm Planck}$. Thus, typically, $\Lambda_{\rm F} \approx \Lambda_{\rm PU} \approx 10^{16 \div 19}$~GeV.

In this section, we will present some general features of PUPC models. The first issue is about choosing the gauge groups. Then, we will show how the SM fermions can be embedded into the PUPC theory, and the scalar sector needed for the symmetry breaking. Finally, we will discuss the conditions under which a walking dynamics can be achieved. In the following two sections we will discuss more gory details about the generation of masses for the third generation first, and then how to extend the theory to the light generations and full flavor structures.

We will start this exploration from the IR end of the spectrum.
It has been shown that only a finite number of gauge-fermion theories can lead to the desired low energy phase~\cite{Barnard:2013zea,Ferretti:2013kya,Vecchi:2015fma}, where both a pNGB Higgs and top PC are achieved.
The latter is due to the presence of light baryonic (spin-1/2) resonances with the SM quantum numbers matching the top field ones.
These theories introduce a new gauge symmetry, called Hyper-Color (HC), with one or two representations for the new hyper-fermions. The strongest constraint on such models comes from the requirement that the gauge dynamics lies outside of the conformal window~\cite{Dietrich:2006cm,Sannino:2009aw,Ryttov:2009yw,Mojaza:2012zd}, i.e. it condenses at low energies and breaks the chiral symmetries in the fermion sector. This requirements leaves only a handful of possibilities~\cite{Ferretti:2016upr}, as it is a strong constraint on the number of fermions and number of hyper-colors. Following the nomenclature of Ref.~\cite{Belyaev:2016ftv}, 12 minimal models have been identified, M1-M12, each characterized by its own gauge group and hyper-fermion representations. As mentioned, such theories lie outside of the conformal window: in order to enter the needed walking phase between $\Lambda_{\rm HC}$ and $\Lambda_{\rm F}$, additional hyper-fermions can be added, with a mass $\sim \Lambda_{\rm HC}$. This IR theory, then, needs to be embedded in the UV PUPC theory, where the HC gauge group is partially unified with the SM one. We will shortly see that this step is non-trivial, and it has consequences for the low energy dynamics, as it can be used to further select the gauge theories in the confined phase.  This selection is crucial in particular for Lattice studies.

The models that achieve the low energy dynamics with top PC resort to HC groups $SO(N)_{\rm HC}$, $SU(N)_{\rm HC}$ and $Sp(N)_{\rm HC}$, with hyper-fermions in the fundamental, spinorial and two-index antisymmetric representations. Following minimality, we decided to unify QCD and HC groups: this is due to the fact that mediators for top PC typically carry QCD charges. As a consequence, we need to embed the hyper-fermion representation and $SU(3)_c$ fundamentals in the same representation of the extended-HC (EHC) group: this is easiest to do for models based on $Sp(N)_{\rm HC}$, like model M8~\cite{Barnard:2013zea,Belyaev:2016ftv}. The reason is that $SO(N)_{\rm HC}$ models always contain the spinorial representation, which is hard to embed together with a fundamental of QCD, while $SU(N)_{\rm HC}$ theories with fundamental tend to inherit the chiral spectrum of the SM in the hyper-fermion sector. While this analysis certainly does not exclude other possibilities, we decided for simplicity to focus on M8, as a template IR model for the first PUPC construction.

The low energy model, therefore, will consist on $Sp(4)_{\rm HC}$ with four hyper-fermions in the fundamental representation: one pair forms a doublet of the gauged $SU(2)_L$ while the other a doublet of the custodial $SU(2)_R$ (the hypercharge corresponds to the diagonal generator). This sector ensures that the pNGB Higgs arises at low energy, and its effect preserves the custodial relation between the $W$ and $Z$ masses. Furthermore, the model needs to include hyper-fermions in the two-index antisymmetric representation in order to obtain top partners in the form of hyper-baryons. The HC and QCD gauge groups are unified as diagonal-subgroups of a $SU(7)_{\rm EHC}$. It is then possible to show that quarks and hyper-fermions in the fundamental can be embedded in fundamentals of $SU(7)_{\rm EHC}$, by suitably choosing the charges under a $U(1)_{\rm E}$, in order to fit the correct hypercharges and cancel gauge anomalies. Leptons here remain as singlets of $SU(7)_{\rm EHC}$, thus they will not receive any contribution to their coupling to hyper-fermions from gauge mediation. This feature, plus the cancellation of anomalies, points towards a unification of quarks--hyper-fermions with leptons, {\it {\`a} la} Pati-Salam~\cite{Pati:1974yy}. Finally, the PUPC gauge group we choose to work with is
\begin{equation}
\mathcal{G}_{\rm TPS} = SU(8)_{\rm PS} \times SU(2)_L \times SU(2)_R\,,
\end{equation}
from which the name of \emph{Techni-Pati-Salam} (TPS) model~\cite{Cacciapaglia:2019dsq}.
The next two questions involve the choices of fermions in the TPS model, which can accommodate for both the chiral SM fermions and the non-chiral hyper-fermions, as well as the choice of scalars, which are responsible for breaking the TPS group down to the SM plus HC gauge symmetries.

\subsection{Fermion embedding}

In the TPS model, both SM fermions and hyper-fermions need to be embedded into representations of the TPS group.
As we will see, the multiplicity and quantum numbers for the hyper-fermions are determined by this choice, thus while we use M8 as a template model, the details
of the IR dynamics will not necessarily be the same.
To indicate the representations, we will use the following notations:
\begin{equation}
\{ {\bf d_{\rm PS}},\ {\bf d}_L,\  {\bf d}_R \}  \Rightarrow \mathcal{G}_{\rm TPS}\,,
\end{equation}
where ${\bf d}_X$ indicates the dimension of the representation under the TPS group $X$, while for the IR quantum numbers we omit the $SU(2)_L$ (as it remains unbroken all the way from the UV to the IR) and use
\begin{equation}
    (d_4,\ d_3)_{Y}  \Rightarrow (Sp(4)_{\rm HC} , SU(3)_c)_{U(1)_Y}\,.
\end{equation}
Details on how the IR gauge groups are embedded in the TPS one in the UV will be presented in the next subsection.

Firstly, for the SM fermions we follow the hint from Pati-Salam~\cite{Pati:1974yy} and we embedded them in a fundamental, $\Omega$, and anti-fundamental, $\Upsilon$,  of $SU(8)_{\rm PS}$, as follows:
\begin{equation}
\Omega = \{ {\bf 8},\ {\bf 2},\ {\bf 1}\} =   \begin{pmatrix} L \\ q_L \\ l_L \end{pmatrix}\,,
\label{eq:Omega}
\end{equation}
\begin{equation}
 \Upsilon = \{ {\bf \bar{8}},\ {\bf 1},\ {\bf 2} \} = \begin{pmatrix} U_d & D_u \\ d^c_R & u^c_R  \\ e^c_R  & \nu^c_R  \end{pmatrix}\,;
\label{eq:Upsilon}
\end{equation}
where all spinors are left-handed Weyl, and the two columns in Eq.~\eqref{eq:Upsilon} explicitly show the two components of the $SU(2)_R$ doublet. The rows follow the $SU(8)_{\rm PS}$ structure, where we embed the IR gauge groups in the following block-diagonal form:
\begin{equation}
   SU(8)_{PS} \Rightarrow \left( \begin{array}{c|c|c}  Sp(4)_{\rm HC} & \phantom{\Big(} &  \\ \cline{1-3}  \phantom{\Big(} &  SU(3)_{c} & \\ \cline{1-3} && \phantom{\Big( xxx}  \end{array} \right)\,.
\end{equation}
One set of $\Omega$ and $\Upsilon$, therefore, contains a complete SM generation
\begin{equation}
\begin{array}{c} \phantom{\Big(} q_L = (1,3)_{1/6}\,, \;\; t_R^c = (1,\bar{3})_{-2/3}\,, \;\; b_R^c = (1,\bar{3})_{1/3}\,, \phantom{\Big(}\\
\phantom{\Big(}  l_L = (1,1)_{-1/2}\,, \;\; e_R^c = (1,1)_{1}\,, \;\; \nu_R^c = (1,1)_0\,, \phantom{\Big(} \end{array}
\end{equation}
including a right-handed neutrino, and the 4 hyper-fermions that generate the pNGB Higgs as a bound state (as in M8)
\begin{equation}
L = (4,1)_0\,, \;\; U_d = (4,1)_{1/2}\,, \;\; D_u = (4,1)_{-1/2}\,.
\end{equation}

Secondly, we need to embed the hyper-fermions in the two-index antisymmetric of $Sp(4)_{\rm HC}$ into the TPS gauge symmetry. The minimal way is to employ antisymmetric representations of $SU(8)_{\rm PS}$: we find convenient and minimal to use the 4-index one, which is a real representation. Other possibilities are discussed in Appendix~\ref{app:irreps}.
The new fermion decomposes as
\begin{equation}
\Xi = \{ {\bf 70},\ {\bf 1},\ {\bf 1} \} = \begin{pmatrix} U_t & \chi & \rho & \eta & \omega \\ D_b & \tilde{\chi} & \tilde{\rho} & \tilde{\eta} & \tilde{\omega} \end{pmatrix}\,,
\end{equation}
where the top row corresponds to fields belonging to a ${\bf 35}$ of $SU(7)_{\rm EHC}$ and the ones in the bottom row to the conjugate representation. Thus, fields in the same column have conjugate quantum numbers.
The components have the following quantum numbers:
\begin{equation}
\begin{array}{c}  \phantom{\Big(} U_t = (4,1)_{-1/2}\,,\;\; \chi = (5,3)_{-1/3}\,, \;\;  \eta = (4,\bar{3})_{-1/6}\,, \phantom{\Big)} \\
\phantom{\Big(} \omega = (1,3)_{-1/3}\,, \;\; \rho = (1,1)_{0}\,. \phantom{\Big)} \end{array}
\end{equation}
We see that the hyper-fermions in the antisymmetric of $Sp(4)_{\rm HC}$ have hypercharge $-1/3$, which does not match the one of M8. As we will see, however, this model set-up allows to construct top partners at low energy. Furthermore, the multiplet $\Xi$ contains two hyperfermions, $U_t$ and $D_b$, with quantum numbers matching $D_u$ and $U_d$ in $\Upsilon$, and a set of hyper-fermions carrying QCD charges, $\eta$/$\tilde{\eta}$. The multiplet also contains fermions that are not charged under the HC group: a vector-like partner of the right-handed bottom, $\omega$/$\tilde{\omega}$, and a singlet $\rho$/$\tilde{\rho}$. All these components may play a role in giving masses to the SM fermions, as we will discuss in the next section.

For now, this should be considered a minimal set of TPS fermions that contain the key players for a correct IR dynamics. The interesting point to remark now is that the TPS embedding fixes the quantum numbers of the hyperfermions and their multiplicity: a set of $\Omega$, $\Upsilon$ and $\Xi$ contains 12 Weyl spinors in the fundamental and 6 Weyl spinors in the antisymmetric of the HC group. Additional HC-singlets are also predicted. As already mentioned, alternative choices are presented in Appendix~\ref{app:irreps}.

\subsection{Scalar sector and TPS symmetry breaking}
\label{sec:scalarpatterns}

\begin{table*}[t!]
\begin{centering}
\begin{tabular}{|c|c|c|}
\hline
  & \multicolumn{2}{c|}{Breaking Pattern} \\
   & $\Psi$--$\Theta$ path & $\Delta$ path \\
\hline
PS breaking \phantom{\Big(} & \multicolumn{2}{c|}{$SU(8)_{\rm PS}\times SU(2)_R \rightarrow SU(7)_{\rm EHC}\times U(1)_E$} \\
\hline
EHC breaking \phantom{\Big(} & $SU(7)_{\rm EHC}\rightarrow SU(4)_{\rm CHC}\times SU(3)_c\times U(1)_X$ & $SU(7)_{\rm EHC}\times U(1)_E\rightarrow SU(4)_{\rm CHC}\times SU(3)_c\times U(1)_Y$ \\
\hline
CHC breaking \phantom{\Big(} & $SU(4)_{\rm CHC}\times U(1)_E \times U(1)_X\rightarrow Sp(4)_{\rm HC}\times U(1)_Y$ &  $SU(4)_{\rm CHC}\rightarrow Sp(4)_{\rm HC}$\\
\hline
\end{tabular}
\caption{Gauge symmetry breaking steps from the UV TPS theory down to the IR HC composite Higgs model. The two paths correspond to two different ways to give VEVs to the scalar fields.
\label{table:gsb}}
\end{centering}
\end{table*}

Various scalar multiplets can accommodate the needed breaking steps between the UV TPS theory and the IR model.
We identified two paths that are of interest for phenomenology, summarized in Table~\ref{table:gsb}, as we will detail in this subsection. We first remark that, besides the gauge symmetry breaking,
scalar fields also play the crucial role of generating masses for the hyper-fermions and mediating PC4F interactions for the SM fermions, and we will see them in action in the next two sections. Here, we limit ourselves to discuss the gauge symmetry breaking patterns.

The breaking of $SU(8)_{\rm PS}$, and splitting of the leptons from quarks/hyper-fermions, can be done in a similar way to the standard Pati-Salam model by introducing
\begin{equation}
\Phi = \{ {\bf 8},\ {\bf 1},\ {\bf 2} \}\,.
\end{equation}
 Once it develops a VEV, which can be aligned as follows\footnote{The two columns correspond to components of $SU(2)_R$.}
\begin{equation}
\langle \Phi \rangle = \frac{v_{\rm PS}^\Phi}{\sqrt{2}} \begin{pmatrix} 0 & 0 \\ \vdots & \vdots \\ 0 & \\ 1 & 0 \end{pmatrix}\,,
\end{equation}
it will break $SU(8)_{\rm PS} \times SU(2)_R \to SU(7)_{\rm EHC} \times U(1)_E$ \cite{Li:1973mq}. The unbroken $U(1)_E$ charge can be expressed as
\begin{equation}
Q_E= T^3_R + \frac{2}{\sqrt{7}} T^8_{\rm PS}\,,
\end{equation}
where $T_R^3$ is the diagonal generator of $SU(2)_R$ and
\begin{equation}
T_{\rm PS}^8 = \frac{1}{4\sqrt{7}} \begin{pmatrix} 1_{7\times 7} &  \\ & -7 \end{pmatrix}\,.
\end{equation}
The fermion multiplets introduced above decompose as
\begin{eqnarray}
\Omega &\Rightarrow & [{\it 7},\,\ {\it 2} ]_{1/14} \oplus [{\it 1},\ {\it 2} ]_{-1/2}\,, \\
\Upsilon & \Rightarrow & [{\it \overline{7}},\ {\it 1} ]_{-1/14 \pm 1/2} \oplus [{\it 1},\ {\it 1}]_{1/2 \pm 1/2}\,, \\
\Xi & \Rightarrow & [{\it 35},\ {\it 1}]_{-2/7} \oplus [{\it \overline{35}},\ {\it 1}]_{2/7}\,,
\end{eqnarray}
where $[{\it SU(7)_{\rm EHC}},\ {\it SU(2)_L}]_{Q_E}$.

The further breaking down to the IR model can follow two paths, which we discuss below.

\subsubsection{The $\Psi$--$\Theta$ path}

The first path requires the following scalar multiplets:
\begin{eqnarray}
\Psi & = & \{ {\bf 63},\ {\bf 1},\ {\bf 1} \}\,, \\
\Theta & = & \{ {\bf 28},\ {\bf 1},\ {\bf 1} \}\,.
\end{eqnarray}

The adjoint $\Psi$ is assumed to develop a VEV proportional to~\cite{Li:1973mq,Elias:1975yd}
\begin{equation}
\langle \Psi \rangle = \frac{v^\Psi_{\rm EHC}}{4} \begin{pmatrix} 1_{4\times 4} & \\ & -1_{4\times 4} \end{pmatrix}\,,
\end{equation}
which, once combined with the $\Phi$ VEV~\cite{Buccella:1979sk,Ruegg:1980gf}, breaks $SU(7)_{\rm EHC} \to SU(4)_{\rm CHC} \times SU(3)_c \times U(1)_X$.
The group $SU(4)_{\rm CHC}$, which we dub complex-HC, contains $Sp(4)_{\rm HC}$, and the would-be hyper-fermions transform as complex representations under the CHC group (see Appendix~\ref{app:irreps} for more details). The unbroken $U(1)_X$ charge corresponds to a diagonal generator of $SU(7)_{\rm EHC}$ that can be expressed in terms of $SU(8)_{\rm PS}$ as
\begin{equation}
Q_X = \frac{1}{42} \begin{pmatrix} 3_{4\times 4} & & \\ & -4_{3\times 3} & \\ & & 0 \end{pmatrix}\,.
\end{equation}
Details about the decomposition of fermion, gauge and scalar multiplets after this step are reported in Appendix~\ref{app:irreps}.

The gauge couplings are matched to the TPS ones as follows:
\begin{eqnarray}
g_{\rm CHC} &=& g_c = g_{\rm PS}\,, \\
g_E &=& \frac{2\sqrt{7} g_R g_{\rm PS}}{\sqrt{4g^2_R + 7g_{\rm PS}^2}}\,, \\
g_X&=&\sqrt{\frac{21}{2}}g_{\rm PS}\,.
\end{eqnarray}
The breaking pattern will also produce massive gauge bosons, among which the most interesting ones are
\begin{equation}
C_\mu = (4,1)_{1/2}\,, \;\; D_\mu = (1,3)_{2/3}\,, \;\; E_\mu = (4,3)_{1/6}\,,
\end{equation}
where the first two form a fundamental of $SU(7)_{\rm EHC}$. As we will see, $E_\mu$ and $C_\mu$ play an important role in mediating PC4F operators, while $D_\mu$ generates four-fermion interactions between quarks and leptons, like in the standard Pati-Salam. Their masses are given by:
\begin{equation}
\begin{array}{c} M^2_E = \frac{g_{\rm PS}^2}{4} (v_{\rm EHC}^\Psi)^2\,, \;\; M^2_C = \frac{g_{\rm PS}^2}{4} (v_{\rm EHC}^\Psi + v_{\rm PS}^\Phi)^2\,, \\
 M_D^2 = \frac{g_{\rm PS}^2}{4} (v_{\rm PS}^\Phi)^2\,, \end{array}
\end{equation}
where we remark that $M_C > M_E$.
For completeness, the spectrum also contains one neutral and one charged singlet deriving from the breaking of $SU(2)_R$, with masses
\begin{equation}
M_{W_R^\pm}^2 = \frac{g_{R}^2}{4} (v_{\rm PS}^\Phi)^2\,, \;\; M_{Z_\Psi}^2 = \frac{4 g_R^2 + 7 g_{\rm PS}^2}{16} (v_{\rm PS}^\Phi)^2\,.
\end{equation}

The next step consists in breaking the CHC group down to $Sp(4)_{\rm HC}$, so that the hyper-fermions can transform under a pseudo-real representation of the HC group.
We will pragmatically assume that this breaking may occur at any energy between $\Lambda_{\rm PS}$ and $\Lambda_{\rm HC}$. Some phenomenological consideration on the relevance of this scale will be presented in the next subsection. To achieve this step, we need a field transforming as a two-index antisymmetric of $SU(4)_{\rm CHC}$, which is naturally contained in $\Theta$, also carrying charges $Q_E = Q_X = 1/7$. A VEV in this component, would also break $U(1)_E \times U(1)_X \to U(1)_Y$, with
\begin{equation}
Y = Q_E - Q_X\,,
\end{equation}
and gauge coupling matching
\begin{equation}
g_Y= \frac{g_E g_X}{\sqrt{g^2_E +g^2_X}}\,, \;\; g_{\rm HC} = g_{\rm CHC}\,.
\end{equation}
The spectrum will now contain two additional gauge bosons, a singlet and $H_\Theta^\mu = (5,1)_0$, with masses
\begin{equation}
M_{H_\Theta}^2 = \frac{g_{\rm CHC}^2}{4} (v_{\rm CHC}^\Theta)^2\,, \;\; M_{Z_\Theta}^2 = \frac{g_E^2 + g_X^2}{4} (v_{\rm CHC}^\Theta)^2\,.
\end{equation}

\subsubsection{The $\Delta$ path}

A second possible path can be achieved by use of a three-index antisymmetric representation
\begin{equation}
\Delta = \{ {\bf 56},\ {\bf 1},\ {\bf 2}\}\,,
\end{equation}
whose VEV can break $SU(8) \to SU(3) \times SU(5)$~\cite{Cummins:1984wt,Adler:2015dba}.
As this VEV also break $U(1)_E$, it needs to transform as an $SU(2)_R$ doublet, with the VEV aligned with the $T_R^3 = -1/2$ component in order to preserve the hypercharge. Thus, together with the $\Phi$ VEV, $\Delta$ can break $SU(7)_{\rm EHC} \times U(1)_E \to SU(4)_{\rm CHC} \times SU(3)_c \times U(1)_Y$.

The matching of the gauge couplings read
\begin{eqnarray}
    g_{\rm CHC} &=& g_c=g_{\rm PS}\,, \\
    g_Y &=&  \frac{g_R g_{\rm PS}}{\sqrt{g^2_R + g^2_R \frac{5}{21}+g^2_{\rm PS}\frac{16}{7}}}\,.
\end{eqnarray}
The spectrum of massive gauge bosons will now read
\begin{equation}
\begin{array}{c} \phantom{\Big(} M_E^2 = \frac{g_{\rm PS}^2}{4} (v_{\rm EHC}^\Delta)^2\,, \;\; M_C^2 = \frac{g_{\rm PS}^2}{4} (v_{\rm PS}^\Phi)^2\,, \phantom{\Big)} \\
 \phantom{\Big(} M_D^2 = \frac{g_{\rm PS}^2}{4} (v_{\rm PS}^\Phi + v_{\rm EHC}^\Delta)^2\,, \;\; M_{W^\pm_R}^2 = \frac{g_{\rm R}^2}{4} (v_{\rm PS}^\Phi + v_{\rm EHC}^\Delta)^2\,; \phantom{\Big)} \end{array}
\end{equation}
plus two massive singlets. We note that $M_C > M_E$ if $v_{\rm PS}^\Phi > v_{\rm EHC}^\Delta$.

Furthermore, the $T_R^3 = 1/2$ component of $\Delta$ contains a component transforming as the two-index antisymmetric of $SU(4)_{\rm CHC}$ with zero hypercharge, thus it can be used to break the CHC symmetry with a VEV $v_{\rm CHC}^\Delta < v_{\rm EHC}^\Delta$. This breaking will simply leave one massive gauge boson, $H_\Delta^\mu = (5,1)_0$, with mass
\begin{equation}
M_{H_\Delta}^2 = \frac{g_{\rm CHC}^2}{4} (v_{\rm CHC}^\Delta)^2\,.
\end{equation}

\subsection{Hypercolor dynamics}
\label{sec:HCdyn}

\begin{figure}[bt]
\includegraphics[width=3in]{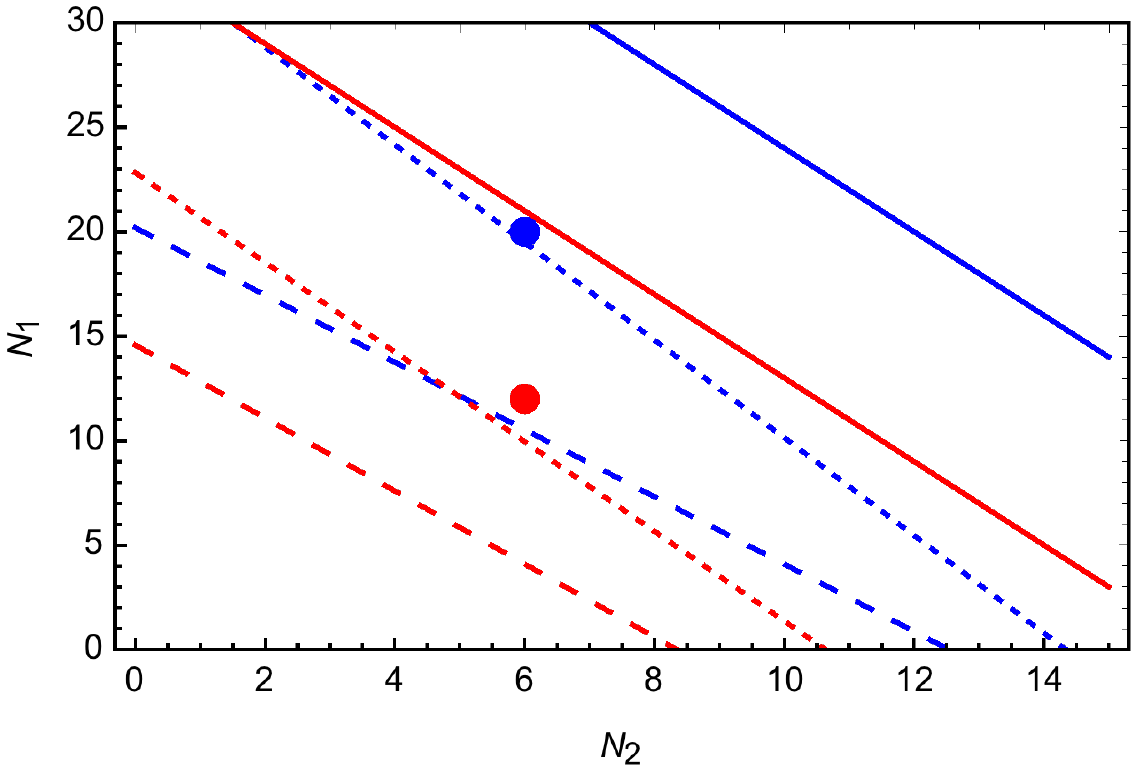}
\caption{Conformal window as a function of the number of Weyl spinors in the fundamental ($N_1$) and antisymmetric ($N_2$) for $Sp(4)_{\rm HC}$ (red) and $SU(4)_{\rm CHC}$ (blue). The solid line indicates where asymptotic freedom is lost, while the dashed and dotted lines indicate the expected lower edge using the PS or SD methods, respectively.\label{fig:confwindow}}
\end{figure}

A key ingredient for any composite Higgs model with top partial compositeness is the presence of a near-conformal ``walking'' dynamics above the condensation scale $\Lambda_{\rm HC}$. This may ensure that the hyper-baryons that couple to the top develop a large anomalous dimensions, which in turn can enhance the top PC couplings at low energy. For this mechanism to have any hope to work, the theory in the walking phase should lie as close as possible to the lower edge of the conformal window, thus being in a strongly coupled regime. Unfortunately, estimating the location of the conformal edge in terms of the fermion multiplicities is subject to many uncertainties, due to the strong coupling. In the following, we will adopt two methods developed in the literature: the Pica-Sannino (PS) all order beta function~\cite{Pica:2010mt}, and the Schwinger-Dyson (SD) equation approach~\cite{Sannino:2009za}.
The former is based on a conjectured all-order beta function that depends on the mass anomalous dimensions of the fermions charged under the running gauge coupling. In the conformal window, the beta function should vanish, while the mass anomalous dimensions are expected to be of order unity. Thus, this provides enough constraints to fix the number of fermions, leading to
\begin{equation}
11 C_2(G) - \sum_r T(r) n_r \left( 3 + \frac{7}{11} \frac{C_2(G)}{C_2(r)} \right) = 0\,,
\end{equation}
where $C_2$ is the Casimir and $T$ the dynkin index of the representation ($G$ indicates the adjoint), while $n_r$ is the number of Weyl fermions in the representation $r$.
The latter method, based on the ladder approximation in the gap equation, makes use of the fact that the anomalous mass dimension at one loop equals 1 at the critical value of the gauge coupling where chiral symmetry is broken. This can be compared to the zero of the beta function, which first appears at two loops, leading to
\begin{equation}
\alpha^\ast = - \frac{4 \pi \beta_0}{\beta_1} = \frac{2 \pi}{3 C_2 (r)}\,.
\end{equation}
As we have two different representations, we will consider the one whose anomalous dimension reaches unity first, i.e. the antisymmetric.

We first apply these methods to a $Sp(4)_{\rm HC}$ theory~\cite{Sannino:2009aw} with $N_1$ Weyl spinors in the fundamental and $N_2$ Weyl spinors in the antisymmetric. The result is shown in Fig.~\ref{fig:confwindow} by the red lines, where the dashed (dotted) correspond to the PS (SD) method. In solid we show the line above which asymptotic freedom is lost. This case is relevant for the TPS model when the CHC breaking occurs at high scale, i.e. before the onset of the walking phase.
The model we presented in this section contains $N_2=6$ degrees of freedom in the antisymmetric representation, coming from the $\Xi$ multiplet. For $N_2=6$, the PC method gives the lower edge starting at $N_1=5$, while for SD it starts at $N_1 = 10$ (while asymptotic freedom is lost for $N_1 = 21$). To compare with a realistic scenario, we recall that one SM generation ($\Omega$ + $\Upsilon$) plus a $\Xi$ contains $N_1 = 12$, which is inside the conformal window (C.f., red dot in Fig.~\ref{fig:confwindow}), and very close to the boundary according to the SD method. Thus, if SD is correct, the model has good chances of being close to the edge and develop large anomalous dimensions. We anticipate that extending to 3 generations would minimally require to add a flavor index to $\Omega$ and $\Upsilon$, raising the number of fundamental hyper-fermions to $N_1 = 20$, which is well too close to the edge of asymptotic freedom loss, where the theory becomes weakly coupled. This simple analysis shows that the hyper-fermions associated to the light generations should not be light, feature that we will exploit in the next sections.

It is also interesting to consider the case where the CHC symmetry is only broken at low energies, after the model enters the walking phase. As the hyper-fermions contained in $\Omega$ and $\Upsilon$ inherit the chiral structure of the SM fermions, they cannot acquire a mass before CHC is broken. Thus, the minimal model with three generations will have $N_1 = 20$. The case of $SU(4)_{\rm CHC}$~\cite{Ryttov:2009yw} is shown in Fig.~\ref{fig:confwindow} in blue, with the same conventions as above: the conformal window edge is expected at $N_1 = 11$ with the PS method, and $N_1 = 20$ with SD (while the asymptotic freedom loss occurs at $N_1 = 32$) The minimal model, represented by the blue dot, is again very close to the SD lower edge of the conformal window.
The case with low scale CHC breaking is therefore also interesting. However, it can only occur if a mechanism that generates a large hierarchy between the VEVs of various scalars is understood. In the following we will focus on the case of high scale CHC breaking, leaving the low scale case for further investigation.

\begin{figure}[bt]
\includegraphics[width=3in]{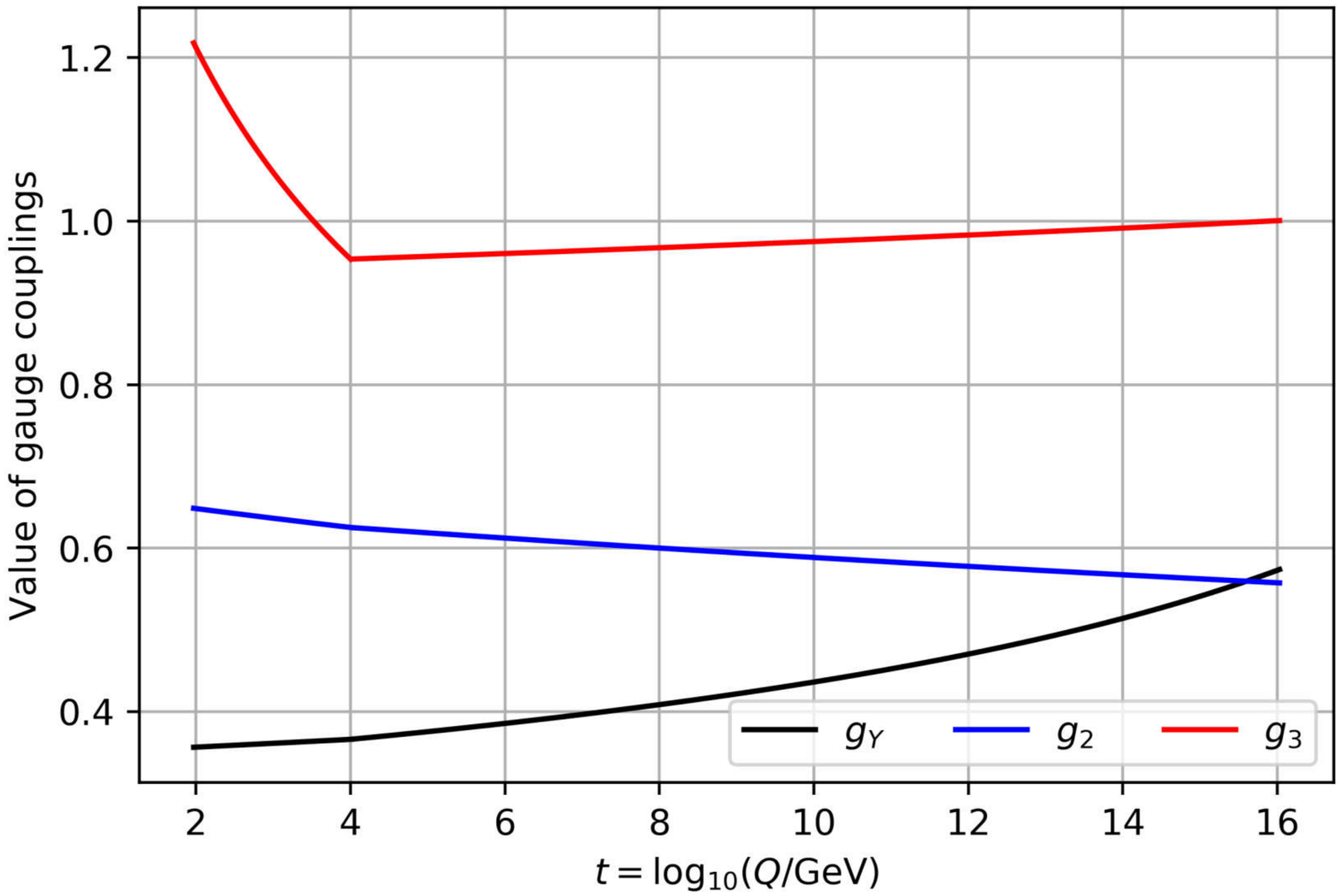}
\caption{Perturbative evolution of SM gauge couplings. Two-loop effects
from SM gauge interactions are taken into account, while HC corrections
are not included.\label{fig:gevolve}}
\end{figure}

The theory we consider in the following, therefore, features the $Sp(4)_{\rm HC}$ dynamics in a walking regime between $\Lambda_{\rm HC}$ and $\Lambda_F$. As a further consistency check, as many fermions are present in this wide energy range, we checked that the running of the SM gauge couplings, $g_3$ for QCD, $g_2$ for $SU(2)_L$ and $g_Y$ for hypercharge do not develop a Landau pole before the $\Lambda_{\rm PU}$ scale. We thus used {\tt PyR@TE}~\cite{Lyonnet:2013dna,Lyonnet:2016xiz} to compute the running where only one generation of hyper-fermions is included (i.e., $N_1 = 12$). The two-loop running is shown in Fig.~\ref{fig:gevolve}, proving that the gauge couplings remain under control. These results are mainly qualitative, as the contribution of the HC gauge coupling, which is strong, has not been included.
There might be concern that $g_3\sim1$ is too perturbative around $\sim 10^{16}\GeV$ where it unifies with
$SU(4)_{CHC}$, so that the resulting $Sp(4)_{HC}$ coupling might spend unacceptably long RG time
in the perturbative regime. However, the ignored HC correction might alter the evolution of $g_3$ so that $SU(4)_{CHC}$ and $SU(3)_C$ unify at some \emph{semi-perturbative} value, which we will assume. Also, above the PU scale, the two $SU(2)$ gauge couplings keep growing as their beta function has lost asymptotic freedom: including 3 generation of $\Omega$ and $\Upsilon$, each has $3\times 8$ Weyl spinors. However, this may be a minor issue, because the Planck scale is close to $\Lambda_{\rm PU}$ by construction, where quantum gravity effects should start to be relevant and may tame the growth of the gauge couplings~\cite{Eichhorn:2018yfc}.

\section{Techni-Pati-Salam for the Third Family}
\label{sec:tps3}

In this section, we will first construct a model that provides masses for one generation of SM fermions, 
namely the third one, as this exercise allows to better illustrate the main properties of the model.
Extension to 3 generations will be presented in the next section. The minimal field content is listed in
Table~\ref{table:sffc}. We add all the scalars discussed in the previous section in order to keep open
both paths of symmetry breaking and also, as we will see, because they all play a crucial role
in generating SM fermion masses.

\begin{table}[tb]
\begin{centering}
\begin{tabular}{|c|c|c|c|c|c|}
\hline
Field & Spin & $SU(8)_{\rm PS}$ & $SU(2)_L$ & $SU(2)_R$  & $Q_G$\\
\hline
$\Phi$ & $0$ & $\textbf{8}$ & $\textbf{1}$ & $\textbf{2}$ & $q$\\
\hline
$\Theta$ & $0$ & $\textbf{28}$ & $\textbf{1}$ & $\textbf{1}$  & $2q$\\
\hline
$\Delta$ & $0$ & $\textbf{56}$ & $\textbf{1}$ & $\textbf{2}$  & $q$\\
\hline
$\Psi$ & $0$ & $\textbf{63}$ & $\textbf{1}$ & $\textbf{1}$  & $0$\\
\hline
$N$ & $1/2$ & $\textbf{1}$ & $\textbf{1}$ & $\textbf{1}$  & $0$\\
\hline
$\Omega$ & $1/2$ & $\textbf{8}$ & $\textbf{2}$ & $\textbf{1}$   & $q$\\
\hline
$\Upsilon$ & $1/2$ & $\overline{\textbf{8}}$ & $\textbf{1}$ & $\textbf{2}$  & $-q$\\
\hline
$\Xi$ & $1/2$ & $\textbf{70}$ & $\textbf{1}$ & $\textbf{1}$ & $0$\\
\hline
\end{tabular}
\caption{Scalar and (left-handed Weyl) fermion field content. 
The last column indicates the global $U(1)_G$ charges, with $q\neq 0$ being an arbitrary normalization
factor.
\label{table:sffc}}
\end{centering}
\end{table}

\subsection{Lagrangian and gauge-mediated PC4F Operators}

The complete Lagrangian of the model, including only renormalizable operators, can be decomposed as
\begin{align}
\mL_{\rm TPS^3}=\mL_G+\mL_F+\mL_S+\mL_Y+\mL_V\,,
\end{align}
where $\mL_G$, $\mL_F$ and $\mL_S$ denote the kinetic terms for gauge, fermion and scalar fields respectively
(including gauge interactions), $\mL_Y$ contains the fermion bare mass terms and Yukawa interactions, while 
$\mL_V=-V(\Phi,\Theta,\Delta,\Psi)$ is the scalar potential term.
For our purposes, the most relevant part is $\mL_Y$, which is given explicitly by:
\begin{align}
\mL_Y & =-\frac{1}{2}\mu_N NN-\frac{1}{2} \mu_\Xi \Xi\Xi-\frac{1}{2} \lambda_\Psi \Xi\Psi\Xi - \left( \lambda_\Phi \Upsilon\Phi N \right. \nonumber \\
& \left. + \lambda_{\Theta L}\Omega\Theta^*\Omega + \lambda_{\Theta R}\Upsilon\Theta\Upsilon
+ \lambda_\Delta \Upsilon\Delta^*\Xi+\text{h.c.} \right)\,,
\label{eqn:lyukawa}
\end{align}
where the first three terms are self-hermitian. In principle, the Yukawas $\lambda_i$ (except $\lambda_\Psi$)
are complex parameters, however one can use arbitrary phase redefinitions of the fermion and scalar fields to make all of
them real, without loss of generality. At this stage, therefore, physical phases can only be contained in the scalar potential
$\mL_V$. The interaction terms in $\mL_Y$ (including the kinetic terms) also leave a global $U(1)_G$ unbroken,
with charges defined in Table~\ref{table:sffc}. Explicit $U(1)_G$-breaking terms may appear in the scalar potential. We assume
minimizing the scalar potential leads to the desired VEV configuration that break the PS, EHC and CHC groups
(see discussion in Sec.~\ref{sec:scalarpatterns}).

The gauge couplings relevant for generating PC4F operators involve only 2 of the massive gauge bosons, deriving
from the PS and EHC breaking: $E_\mu = ({\bf 4}, {\bf 3})_{1/6}$ and $C_\mu = ({\bf 4}, {\bf 1})_{1/2}$ . 
 Their couplings read~\footnote{According to our 
normalization and sign convention, the covariant derivative of a fermion $\psi_i$
in the fundamental of $SU(8)_{PS}$ is written as $D_\mu\psi_i=\partial_\mu\psi_i-i\frac{g_{PS}}{\sqrt{2}}
W^j_{\mu i}\psi_j$ with $i,j$ being $SU(8)$ indices. The same convention is used for $SU(7)_{EHC}$.}
\begin{equation}
\mL_F \supset \frac{g_{\rm PS}}{\sqrt{2}}\ C_\mu J_C^\mu + \frac{g_{\rm EHC}}{\sqrt{2}}\ E_\mu J_E^\mu + \text{h.c.}\,,
\end{equation}
where $g_{\rm PS}$ and $g_{\rm EHC}$ are the gauge couplings of $SU(8)_{\rm PS}$ and
$SU(7)_{\rm EHC}$ respectively, with $g_{\rm EHC} \approx g_{\rm PS}$ if the breaking of the two symmetries
is happening at closeby scales.
The two currents read:
\begin{widetext}
\begin{multline}
J_E^\mu  =  \bar{q_L}\bar{\sigma}^\mu L^3-\bar{D}_u^3\bar{\sigma}^\mu t_R^c
-\bar{U}_d^3\bar{\sigma}^\mu b_R^c
 + \frac{1}{2} \left( \bar{\chi} \bar{\sigma}^\mu U_t -\bar{D}_b\bar{\sigma}^\mu \tilde{\chi} \right)
- \left( \bar{\eta}\bar{\sigma}^\mu\chi
-\bar{\tilde{\chi}}\bar{\sigma}^\mu\tilde{\eta} \right) \\
- \left( \bar{\eta} \bar{\sigma}^\mu \omega - \bar{\tilde{\omega}} \bar{\sigma}^\mu \tilde{\eta} \right)
 + \frac{1}{2} \left( \bar{\rho} \bar{\sigma}^\mu \eta - \bar{\tilde{\eta}} \bar{\sigma}^\mu \tilde{\rho} \right) 
+ \frac{1}{2} \left( \bar{\omega} \bar{\sigma}^\mu U_t - \bar{D}_b \bar{\sigma}^\mu \tilde{\omega} \right)\,,
\label{eq:JE}
\end{multline}
\begin{multline}
J_C^\mu  =\bar{L}^3\bar{\sigma}^\mu l_L
- \bar{\nu}_{\tau R}^c\bar{\sigma}^\mu D_u^3 - \bar{\tau}_R^c\bar{\sigma}^\mu U_d^3 
 -\frac{1}{2} \left( \bar{\tilde{\eta}}\bar{\sigma}^\mu \chi+\bar{\tilde{\chi}}\bar{\sigma}^\mu \eta \right) 
-\frac{1}{2} \left( \bar{\tilde{\eta}} \bar{\sigma}^\mu \omega + \bar{\tilde{\omega}} \bar{\sigma}^\mu \eta \right)
-\frac{1}{6} \left( \bar{\tilde{\rho}} \bar{\sigma} U_t + \bar{D}_b \bar{\sigma} \rho \right)\,.
\label{eq:JC}
\end{multline}
\end{widetext}
By integrating out the two vector mediators, we obtain the following four fermion operators, linear in the
SM fields: 
\begin{widetext}
\begin{multline}
\mathcal{L}_{\rm PC4F} \supset  - \frac{g_{\rm EHC}^2}{2 M_E^2} \ 
\left( \bar{L}^3\bar{\sigma}^\mu q_L
-\bar{t}_R^c \bar{\sigma}^\mu D_u^3
-\bar{b}_R^c \bar{\sigma}^\mu U_d^3 \right) 
\left( \frac{1}{2}\bar{\chi} \bar{\sigma}_\mu U_t
-\frac{1}{2}\bar{D}_b \bar{\sigma}_\mu  \tilde{\chi}
-\bar{\eta} \bar{\sigma}_\mu \chi
+\bar{\tilde{\chi}} \bar{\sigma}_\mu  \tilde{\eta} \right) \\
- \frac{g_{\rm PS}^2}{2 M_C^2} \
\left( \bar{L}^3\bar{\sigma}^\mu l_L
-\bar{\nu}_{\tau R}^c \bar{\sigma}^\mu D_u^3
-\bar{\tau}_R^c\bar{\sigma}^\mu U_d^3\right) 
\left( -\frac{1}{2} \bar{\chi} \bar{\sigma}_\mu  \tilde{\eta}
-\frac{1}{2}\bar{\eta} \bar{\sigma}_\mu \tilde{\chi}
\right)\,.
\label{eqn:vmpc4f}
\end{multline}
\end{widetext}
The interesting property of Eq.~\eqref{eqn:vmpc4f} is that all quark operators are mediated by $E_\mu = (\textbf{4},\textbf{3})_{1/6}$,
 which becomes massive from the EHC breaking, while all lepton operators are mediated by 
 $C_\mu = (\textbf{4},\textbf{1})_{1/2}$, which becomes massive from the PS breaking.
 The mass hierarchy between leptons and quarks could, therefore, be explained by a hierarchy
 in the masses of the mediators if $M_C > M_E$ (see Sec.~\ref{sec:scalarpatterns}). Furthermore, lepton operators
 always involve the QCD-colored hyper-fermions $\eta$--$\tilde{\eta}$, while the quark ones also involve
 the QCD-singlets $U_t$ and $D_b$. 

It is remarkable that our PUPC approach allows to generate appropriate PC4F operators for all SM quarks
from gauge interactions, however  there is no distinction between fermions in the same weak isospin
multiplet. In other words, the gauge interactions themselves cannot distinguish between top-bottom,
nor between tau-neutrino. Such mass splittings, which need violation of $SU(2)_R$, naturally receive
contributions in our model: from scalar mediated PC4F operators, from the masses of the involved hyper-fermions,
and, in the case of the neutrino, from mixing with the singlet $N$ via $\lambda_\Phi$. These effects
are discussed in the following sub-sections.

\begin{table*}[tb]
\begin{centering}
{\footnotesize \begin{tabular}{|c|c|c|c|c|c|c|c|c|c|c|c|c|c|c|}
\hline
 & \multicolumn{7}{c|}{1 SM field} & \multicolumn{7}{c|}{0 SM field} \\ \hline
\phantom{$\Big($} $\varphi_i$ & $({\bf 4}, {\bf 1})_{-\frac{1}{2}}$ & $({\bf 4}, {\bf 3})_{\frac{1}{6}}$ & $({\bf 4}, {\bf 3})_{-\frac{5}{6}}$ & $({\bf 5}, {\bf 1})_{0}$ & $({\bf 5}, {\bf 1})_{-1}$ & $({\bf 5}, {\bf 3})_{\frac{2}{3}}$ & $({\bf 5}, {\bf 3})_{-\frac{1}{3}}$ &   $({\bf 4}, {\bf 1})_{\frac{1}{2}}$ & $({\bf 4}, {\bf 3})_{\frac{1}{6}}$ & $({\bf 4}, {\bf 3})_{-\frac{5}{6}}$ & $({\bf 5}, {\bf 1})_{0}$ & $({\bf 5}, {\bf 1})_{-1}$ & $({\bf 5}, {\bf 3})_{\frac{2}{3}}$ & $({\bf 5}, {\bf 3})_{-\frac{1}{3}}$ \\ 
\hline \hline
$\Omega \Theta^\ast \Omega$ & $(L^3 l_L)$ & $(L^3 q_L)$ & - & - & - & - & - & - & - & - & $(L^3 L^3)$ & - & - & - \\
\hline
\multirow{2}{*}{$\Upsilon \Theta \Upsilon$} & \textcolor{red}{$(U_d^3 \nu_R^c)$} & \textcolor{red}{$(U_d^3 t_R^c)$}  & - & - & - & - & - & 
- & - & - & \multirow{2}{*}{$(U_d^3 D_u^3)$} & - & - & - \\
 &  \textcolor{red}{$(D_u^3 \tau_R^c)$} & \textcolor{red}{$(D_u^3 b_R^c)$}  & - & - & - & - & - & 
- & - & - &  & - & - & - \\ 
\hline
\multirow{4}{*}{$\Xi \Psi \Xi$} & \multirow{4}{*}{-} & \multirow{4}{*}{-}  & \multirow{4}{*}{-}  & \multirow{4}{*}{-}  & \multirow{4}{*}{-}  & \multirow{4}{*}{-}  & \multirow{4}{*}{-}  &
 & $(\chi D_b)$ & \multirow{4}{*}{-} & \multirow{2}{*}{$(U_t D_b)$} & \multirow{4}{*}{-} & \multirow{4}{*}{-} & \multirow{4}{*}{-}  \\
& & & & & & & & 
& \textcolor{red}{$(U_t \tilde{\chi})$} & & & & & \\
& & & & & & & & 
$(\chi \eta)$ & $(\eta \tilde{\chi})$ & & \multirow{2}{*}{$(\eta \tilde{\eta})$} & & & \\
& & & & & & & & 
\textcolor{red}{$(\tilde{\chi} \tilde{\eta})$} &  \textcolor{red}{$(\chi \tilde{\eta})$} & & & & & \\
\hline \hline
\multirow{4}{*}{$\Upsilon \Delta^\ast \Xi$} & $(U_t \nu_R^c)$ & \multirow{2}{*}{\textcolor{red}{$(D_b t_R^c)$}} & \multirow{2}{*}{\textcolor{red}{$(D_b b_R^c)$}} & & & & & 
\multirow{4}{*}{-} & \multirow{2}{*}{$(\chi U_d^3)$} & \multirow{2}{*}{$(\chi D_u^3)$} &  \multirow{4}{*}{$(U_t U_d^3)$} &  \multirow{4}{*}{$(U_t D_u^3)$} &  \multirow{4}{*}{$(\tilde{\eta} U_d^3)$} &   \multirow{4}{*}{$(\tilde{\eta} D_u^3)$} \\
& \textcolor{red}{$(U_t \tau_R^c)$} & & & & & & & 
& & & & & & \\
& $(\tilde{\eta} t_R^c)$ & $(\eta b_R^c)$ & $(\eta t_R^c)$ & $(\chi b_R^c)$ & $(\chi t_R^c)$ & $(\tilde{\chi} b_R^c)$ & $(\tilde{\chi} t_R^c)$ &
& \multirow{2}{*}{\textcolor{red}{$(\tilde{\chi} D_u^3)$}} & \multirow{2}{*}{\textcolor{red}{$(\tilde{\chi} U_d^3)$}} & &  & & \\
& \textcolor{red}{$(\tilde{\eta} b_R^c)$} &  \textcolor{red}{$(\eta \nu_R^c)$} &  \textcolor{red}{$(\eta \tau_R^c)$} &  & & $(\chi \tau_R^c)$ & $(\chi \nu_R^c)$ & 
& & & & & & \\
\hline
\end{tabular}}
\caption{\label{table:PC4Fscalar} Scalar mediators $\varphi_i$ (quantum numbers listed in the top row), with the fermion bilinears they couple with. The rows correspond to  different
Yukawa interactions from $\mL_Y$. The fermion bilinears in red couple to the conjugate scalar, $\varphi_i^\ast$.}
\end{centering}
\end{table*}

\subsection{Scalar mediated PC4F operators}

The Yukawa couplings in $\mL_Y$, Eq.~\eqref{eqn:lyukawa}, allow for many scalar components
to mediate PC4F operators. All the relevant combinations are listed in Table~\ref{table:PC4Fscalar},
where we have identified 7 distinct mediators, whose quantum numbers are listed in the top row.
The rows correspond to different Yukawa couplings, while the left block ``1 SM field'' contains fermion bilinears
containing one SM field and the right one ``0 SM field''  bilinears involves only hyper-fermions. The PC4F operators
can thus be constructed by coupling one fermion bilinear from the left block with one from the
right block, if they have matching quantum numbers. If they belong to different Yukawa couplings, 
the resulting operator can only be generated if the components in the two scalar multiplets mix.
As an example, the mediators $\varphi_4 = ({\bf 5}, {\bf 1})_0$ and $\varphi_5 = ({\bf 5}, {\bf 1})_{-1}$,
components of $\Delta$, will generate the following PC4F operators for right-handed top and bottom:
\begin{equation}
\mathcal{L}_{\rm PC4F} \supset - \frac{\lambda_\Delta^2}{M_{\varphi_4}^2} c_4\ (\overline{U}_t \overline{U}_d^3) (\chi b_R^c) - 
 \frac{\lambda_\Delta^2}{M_{\varphi_5}^2} c_5\ (\overline{U}_t \overline{D}_u^3) (\chi t_R^c)\,,
 \label{eq:scalarPC4F_1}
\end{equation}
where $c_{4,5}$ are group theory factors. This example illustrates how a mass splitting between
top and bottom could arise if the above couplings are dominant, and there exist a significant mass
difference between the two scalar mediators. Scalar-mediated PC4F operators are subject to a larger degree of
arbitrariness compared to vector-mediated PC4F operators, because their strengths are determined by
the non-universal Yukawa couplings, and masses and mixing of scalar components controlled by details of
the scalar potential. Nevertheless, they are also generated automatically from the renormalizable
Lagrangian, rather than being put in by hand.

The main ingredients that determine the relevance of scalar mediated PC4F operators are the following:
\begin{itemize}

\item[-] the masses and mixing pattern of the scalars. 

\item[-] the size of the Yukawa couplings. As we will see in the next section, the masses of the
hyper-fermions also depend on some of these Yukawas. To keep some hyper-fermions light, therefore,
a number of Yukawas need to be small, thus also being ineffective in generating sizable PC4F
operators.

\end{itemize}

In the next 3 subsections, we will discuss the impact of the Yukawa couplings on the hyper-fermion masses,
and list the concrete ways the model allows to generate the top-bottom mass hierarchy and small
neutrino masses.

\subsection{Hyper-fermion masses} \label{sec:hypfmass}

Hyper-fermion masses play an important role in determining the properties of the model. Firstly, the
low-energy global symmetry pattern is determined by the number of hyper-fermions that are lighter than
the hypercolor condensation scale $\Lambda_{\rm HC}\sim 10\TeV$. Secondly, whether the HC dynamics
 enters a strongly-coupled near-conformal regime above $\Lambda_{\rm HC}$ depends on the additional 
 hyper-fermions that have a mass between $\Lambda_{\rm HC}$ and $\Lambda_{\rm EHC}$, as discussed 
 in Sec.~\ref{sec:HCdyn}. Thirdly, the mass of the hyper-fermions participating in the PC4F operators determines 
 the masses of the corresponding SM fermions: assuming that the dominant contribution is coming from
 local insertions of the PC4F operators, the SM fermion mass is proportional to the corresponding Fourier-transformed
two-point hyperbaryon correlator at zero momentum~\cite{Golterman:2015zwa}. When one of the
participating hyper-fermions is heavier than $\Lambda_{\rm HC}$, the
correlator is expected to be suppressed by some power of the hyper-fermion mass, as it can be analyzed
via the Shifman-Vainshtein-Zakharov (SVZ) expansion~\cite{Shifman:1978bx,Shifman:1978by}.

Let's start the discussion with the hyper-fermions $\chi$--$\tilde{\chi}$ and $\eta$--$\tilde{\eta}$: they enter 
in all PC4F operators for quarks and leptons, thus they cannot be too heavy. In particular, as all quark
operators, both from gauge and scalar mediation, contain $\chi$ or $\tilde{\chi}$, while all fermion 
operators contain $\eta$ or $\tilde{\eta}$, in order to obtain a large enough top and tau masses
it would be optimal to have $M_\chi \leq \Lambda_{\rm HC}$ and $M_\eta \leq \mathcal{O} (10) \times \Lambda_{\rm HC}$.
Furthermore, a hierarchy $M_\chi < M_\eta$ could explain why leptons are lighter than quarks.
These hyper-fermion masses receive contributions only from the $\Xi$ mass term and from the Yukawa $\lambda_\Psi$ via
the $\Psi$ VEV, resulting in the following terms
\begin{align}
\mL_Y & \supset -\mu_0 U_t D_b -(\mu_0-5\mu_1)(\tilde{\chi}\chi+\tilde{\omega}\omega) \nonumber \\
& -(\mu_0+2\mu_1)\tilde{\eta}\eta-\mu_0\tilde{\rho}{\rho}+\text{h.c.}
\label{eqn:ly1}
\end{align}
where
\begin{equation}
\mu_0 \propto \mu_\Xi\,, \qquad \mu_1 \propto \lambda_\Psi v^\Psi_{\rm EHC}\,.
\end{equation}
Note that, as expected, $\mu_0$ is a universal term for all components of $\Xi$, while
$\mu_1$ only contributes to a sub-set of them.
Thus, we can identify
\begin{equation}
M_\chi = | \mu_0 - 5 \mu_1 |\,, \qquad M_\eta = |\mu_0 + 2 \mu_1 |\,,
\end{equation}
while the masses of the other components receive additional contributions via mixing, as we
will discuss below. The desired hierarchy $M_\chi < M_\eta$ is thus achieved for $0 < \mu_1 < \frac{2}{3} \mu_0$,
where we have assumed $\mu_0 > 0$ without loss of generality. The value of the parameter $\mu_0$, which
contributes to the mass of the singlet $\rho$--$\tilde{\rho}$ and of the hyper-fermions $U_t$--$D_b$, is related
to the two masses by the inequality
\begin{align}
\mu_0\leq\frac{2}{7}M_\chi+\frac{5}{7}M_\eta\,,
\end{align}
implying that is tends to be smaller than the two masses.
An important lesson we can take from this analysis is that, barring fine cancellations,  $\mu_0, \mu_1 \ll \Lambda_{EHC}$, which implies that
the Yukawa $\lambda_\Psi$ needs to be very small. This is technically natural, however it has an important consequence
on the scalar mediated PC4F operators: the ones stemming from $\Xi \Psi \Xi$ (see Table~\ref{table:PC4Fscalar}) are highly suppressed.

We can now discuss the masses of the QCD-singlet hyper-fermions, $L^3$, $D_u^3$, $U_d^3$, $U_t$ and $D_b$, which are relevant for 
generating the composite Higgs at low energy. The pNGB Higgs, in fact, is a bound state of $L^3$ and one of the weak iso-singlets: this
implies that one needs $L^3$ and one set of the iso-singlets to be much lighter that $\Lambda_{\rm HC}$. 
While the $\Xi$ components $U_t$--$D_b$ receive a mass from Eq.~\eqref{eqn:ly1}, the other hyper-fermions
receive a mass via the $\Theta$-VEV as follows:
\begin{equation}
\mL_Y \supset - \mu_L L^3 L^3 - \mu_R U_d^3 D_u^3\,,
\end{equation}
where
\begin{equation}
\mu_L = \lambda_{\Theta L} v^\Theta_{\rm CHC}\,, \qquad \mu_R = \lambda_{\Theta R} v^\Theta_{\rm CHC}\,.
\end{equation}
For the iso-doublet, this is the only contribution to the mass, so that $M_L = \mu_L$. To keep this mass
small, there are three possibilities: a) $\lambda_{\Theta L} \ll 1$ and $v^\Theta_{\rm CHC} \gg \Lambda_{\rm HC}$, thus
the scalar-mediated PC4F operators cannot receive contributions from  $\Omega \Theta^\ast \Omega$; b) $\lambda_{\Theta L} \lesssim 1$
and $v^\Theta_{\rm CHC} \geq \Lambda_{\rm HC}$; c) $v^\Theta_{\rm CHC} = 0$. In the last two cases the Yukawa could
give sizable contributions to scalar-mediated PC4F operators. 
In the case of the iso-singlets, mixing terms are also generated in the presence of VEVs for $\Delta$, in the form
\begin{align}
\mL_Y & \supset -\mu_{\Delta 1} \left( D_u^3 D_b - \nu_{\tau R}^{c} \rho \right) \nonumber \\
& -\mu_{\Delta 2} \left( U_d^3 U_t + \sqrt{2} b_R^{c}\omega \right)  
+\text{h.c.}
\label{eqn:ly2}
\end{align}
where
\begin{align}
\mu_{\Delta 1} = \lambda_\Delta v^\Delta_{\rm EHC}\,, \qquad \mu_{\Delta 2} = \lambda_\Delta v^\Delta_{\rm CHC}\,.
\end{align}
Note that these two terms also induce a mixing of $\rho$ with the neutrinos, and of $\omega$ with the right-handed bottom.
We will come back to their effect in the next two subsections.
In the hyper-fermion sector, this leads to the following mass matrix:
\begin{align}
\mL_Y & \supset
-\begin{pmatrix}
U_d^3 & D_b
\end{pmatrix}
\begin{pmatrix}
\mu_{R} & \mu_{\Delta 2} \\
\mu_{\Delta 1} & \mu_0
\end{pmatrix}
\begin{pmatrix}
D_u^3 \\ U_t
\end{pmatrix}+\text{h.c.}\,,
\label{eq:fr}
\end{align}
which has eigenvalues
\begin{align}
M_{R1,2}^2 & = \frac{1}{2} \left( \tilde{\mu}^2 \mp \sqrt{\tilde{\mu}^4 - 4 (\mu_0 \mu_R - \mu_{\Delta 1} \mu_{\Delta 2})^2} \right)\,, \nonumber \\
& \mbox{with}\;\; \tilde{\mu}^2 = \mu_0^2 + \mu_R^2 + \mu_{\Delta 1}^2 + \mu_{\Delta 2}^2\,.
\end{align}
We see that one can achieve at least one small mass eigenvalue if either all $\mu$'s are small, or
\begin{align}
2 (\mu_0 \mu_R - \mu_{\Delta 1} \mu_{\Delta 2}) \ll \tilde{\mu}^2\,.
\end{align}
Seen the constraints on $\mu_0$ coming from the $\chi$ and $\eta$ masses, the latter
condition may be achieved for
\begin{align}
\mbox{a)} \phantom{xxx} & \mu_R \ll \mu_0\,, \;\; \mu_{\Delta 1} \mu_{\Delta 2} \ll \mu_0^2\,, \nonumber \\
\Rightarrow &  M_{R1} \approx \left| \mu_R - \frac{\mu_{\Delta 1} \mu_{\Delta 2}}{\mu_0} \right|\,, \;\; M_{R2} \approx \mu_0\,;
\end{align}
or
\begin{align}
\mbox{b)}  \phantom{xxx} & \mu_0 \ll \mu_R\,, \;\; \mu_{\Delta 1} \mu_{\Delta 2} \ll \mu_R^2\,, \nonumber \\
\Rightarrow &  M_{R1} \approx \left| \mu_0 - \frac{\mu_{\Delta 1} \mu_{\Delta 2}}{\mu_R} \right|\,, \;\; M_{R2} \approx \mu_R\,.
\label{eq:isosingletmasses}
\end{align}
In the latter case, if $\mu_R \geq \Lambda_{\rm HC}$, one could have that only one mass eigenstate
is below the condensation scale, while in the former typically both are light. One can see, therefore, that
the masses have a crucial impact on the low energy dynamics of the theory by influencing the
global coset that determines the properties of the composite Higgs:
\begin{align}
M_{R2} \geq \Lambda_{\rm HC} \Leftrightarrow \frac{SU(4)}{Sp(4)} \;\; \mbox{\cite{Galloway:2010bp,Cacciapaglia:2014uja}}\,, \nonumber \\
M_{R2} \ll \Lambda_{\rm HC} \Leftrightarrow \frac{SU(6)}{Sp(6)}\;\;  \mbox{\cite{Cai:2018tet}}\,. \nonumber
\end{align}
We also remark that, keeping $\mu_{\Delta 1} \mu_{\Delta 2}$ small would imply either $\lambda_{\Delta} \ll 1$,
or a large hierarchy between the VEVs, $v^\Delta_{\rm CHC} \ll v^\Delta_{\rm EHC}$, with the extreme
case $v^\Delta_{\rm CHC} = 0$. These various possibilities have an important impact on the scalar PC4F
sector, by determining which terms can be sizable and which ones are always suppressed. The implications
for the masses of leptons and quarks will be discussed in the following two subsections.

We recall that the patterns of hyper-fermion masses depend crucially on the pattern of VEVs that break
the TPS group down to the low energy theory.  In this discussion we work under the assumption that the 
desired vacuum misalignment and EWSB can be achieved,
leaving a detailed study of the vacuum misalignment mechanism to future work~\cite{Giacomo:2019ehd}.

To conclude, we would like to recap the main findings in two special cases of VEV patterns, following
the discussion in Sec.~\ref{sec:scalarpatterns}.

\begin{itemize}

\item[A)] $\langle \Delta \rangle = 0$. In this case, the EHC breaking is due to $v^\Psi_{\rm EHC}$,
while $v^\Theta_{\rm CHC}$ breaks $SU(4)_{\rm CHC}$ down to $Sp(4)_{\rm HC}$. The mixing
terms between iso-singlet hyper-fermions vanish, so that we have a simple mass pattern:
\begin{align}
M_L = \mu_L\,, \;\; M_{R1} = \mbox{min} \{ \mu_R, \mu_0 \}\,, \nonumber \\
 M_{R2} = \mbox{max} \{ \mu_R, \mu_0 \}\,. 
\end{align}
Furthermore, the HC-singlets $\omega$ and $\rho$ do not mix and have masses
\begin{align}
M_\omega = M_\chi\,, \;\; M_\rho = \mu_0\,.
\end{align}
The only large Yukawa is therefore $\lambda_\Delta$, which is responsible for generating scalar PC4F
operators (one could also have sizable $\lambda_{\Theta L/R}$ if $v^\Theta_{\rm CHC} \approx \Lambda_{\rm HC}$).
Note that keeping $M_\chi$ below $\Lambda_{\rm HC}$ requires small $\mu_1$,
where the hierarchy $M_\chi < M_\eta$ can be kept for $0 < \mu_1 < 2/3\ \mu_0$.

\item[B)] $\langle \Theta \rangle = \langle \Psi \rangle = 0$. In this case, both EHC and CHC breaking
is due to VEVs of the field $\Delta$. As $\mu_1 = \mu_L = \mu_R = 0$, we have
\begin{align}
M_\chi = M_\eta = \mu_0\,, \;\; M_L = 0\,,
\end{align}
while the iso-singlet masses are given by Eq.\eqref{eq:isosingletmasses} with $\mu_R = 0$. At least 
one light eigenstate can be achieved by keeping the mixing terms small, thus requiring $\lambda_\Delta \ll 1$
(and the corresponding Yukawa ineffective in generating scalar PC4F operators).

\end{itemize}

\subsection{Top-Bottom Mass Splitting}

The SM features a large hierarchy between top and bottom masses, with $m_t/m_b\sim 60$ at the weak scale. 
In the TPS model, the top-bottom mass splitting must be traced back to
spontaneous $SU(2)_R$-breaking. We identified three effects that may explain this feature,
which we analyse in detail below.

Firstly, we noted that gauge mediators as well as scalar mediators from the $\Upsilon \Theta \Upsilon$ Yukawa
cannot be used as they contain both $b_R^c$ and $t_R^c$.  However, scalar-mediated PC4F operators constructed 
from $\Upsilon\Delta\Xi$ involve mediators that differ in type and
properties for $t_R^c$ and $b_R^c$, as it can be seen in Table~\ref{table:PC4Fscalar}.  Thus, a split between top and 
bottom can simply arise from a difference in mass between the two mediators. One example shown in Eq.~\eqref{eq:scalarPC4F_1}
involves $\varphi_4 = ({\bf 5}, {\bf 1})_0$ and $\varphi_5 = ({\bf 5}, {\bf 1})_{-1}$. Another example
involves $\varphi_2 = ({\bf 4}, {\bf 3})_{1/6}$ and $\varphi_3 = ({\bf 4}, {\bf 3})_{-5/6}$. In both cases, the scalar mass difference
breaks $SU(2)_R$, and a sizable coefficient can arise from a large $\lambda_\Delta$, allowed for vanishing $\Delta$ 
VEV.~\footnote{In a less minimal model, this effect could also arise in presence of multiple $\Delta$-multiplets. }
Another source of mass split lies in the fact that the quantum number $(\tbf{5},\tbf{1})_0$ has more ways of pairing compared to $(\tbf{5},\tbf{1})_{-1}$
since it also appears in Yukawa terms other than $\Upsilon\Delta\Xi$. Note this is not incompatible with the fact that
the Yukawa Lagrangian explicitly preserves $SU(2)_R$ which is a gauge symmetry. The reason is that the required mixing between
scalar components with quantum number $(\tbf{5},\tbf{1})_0$ can only occur if there exists spontaneous $SU(2)_R$-breaking
from the scalar potential. Let us also note that this mechanism does not lead to a prediction of the top-bottom
mass splitting, nor a prediction of which quark is heavier, because these properties sensitively depend on details of the scalar potential.

Secondly, a differentiation of top and bottom may come from the mixing in the iso-singlet hyper-fermion sector, given by
Eq.~\eqref{eq:fr}. This opens the possibility that the top has a larger coupling to the lighter mass eigenstate, while
the bottom dominantly couples to the heavier one, thus having its mass suppressed. To be more specific, we can
analyse the case of dominant gauge mediation: from Eq.~\eqref{eq:JE} we see that $t_R^c$ couples to $D_u^3$, while 
$b_R^c$ to $U_d^3$. As the mixing angles for the pairs $D_u^3$--$U_t$ and $U_d^3$--$D_b$ are different if
$\mu_{\Delta 1} \neq \mu_{\Delta 2}$,  one can easily generate hierarchical mixing angles. For instance, for 
$\mu_R = 0$ (achieved if $\langle \Theta \rangle = 0$) the mixing relevant for the top is proportional to $\mu_{\Delta 2}$,
while the one for the bottom to $\mu_{\Delta 1}$. As 
\begin{align}
\frac{\mu_{\Delta 1}}{\mu_{\Delta 2}} \propto \frac{v^\Delta_{\rm EHC}}{v^\Delta_{\rm CHC}} > 1\,,
\end{align}
a larger mixing angle for the bottom is assured. Another interesting possibility is that both iso-singlet hyper-fermions
remain light, in which case the theory features two Higgs doublets in the IR, and the mass hierarchy may be due to
the distribution of the EW VEV on the two doublets~\cite{Rosenlyst:2020znn}, as in traditional 2HDM~\cite{Branco:2011iw}.

Thirdly, the most interesting mechanism sprouts from the mixing between $b_R^c$ and $\omega$,
see Eq.~\eqref{eqn:ly2}. As no such term exists for the top quark, this mixing leads to a suppression of the bottom mass.
The complete mass term reads
\begin{align}
\mL_Y \supset - \omega \left( (\mu_0 - 5 \mu_1) \tilde{\omega} + \sqrt{2} \mu_{\Delta 2} b_R^c \right) + \text{h.c.}
\end{align}
Thus we can define mass eigenstates as
\begin{align}
B_L = \omega\,, \;\; & B_R^c = \cos \alpha_b\ \tilde{\omega} + \sin \alpha_b\ b_R^c\,, \nonumber \\
& \tilde{b}_R^c = \cos \alpha_b\ b_R^c - \sin \alpha_b\ \tilde{\omega}\,,
\end{align}
where
\begin{align}
\tan \alpha_b & = \text{sign} (\mu_0 - 5 \mu_1)\ \frac{\sqrt{2} \mu_{\Delta 2}}{M_\chi}\,, \\
M_B & = \sqrt{M_\chi^2 + 2 \mu_{\Delta 2}^2} \geq M_\chi\,,
\end{align}
while $\tilde{b}_R^c$ can be identified with the (massless) right-handed bottom. In the case of gauge mediation,
the current in Eq.~\eqref{eq:JE} can be re-written as
\begin{align}
J_E^\mu \supset \left(- \cos (\alpha_b) \bar{U}_d^3 - \frac{1}{2} \sin (\alpha_b) \bar{D}_b \right) \bar{\sigma}^\mu \tilde{b}_R^c + \dots
\end{align}
Combined with the mixing between $U_d^3$--$D_b$, this could lead to a suppressed coupling of the right-handed bottom to 
the PC4F operators.

It is also instructive to study a case where an effective mass term for the bottom is induced in the form $- \mu_b b_L b_R^c$.
The mixing with $\omega$ will therefore appear as:
\begin{align}
\mL_{b\omega}=-\begin{pmatrix}
b_L & \omega
\end{pmatrix}
\begin{pmatrix}
m_{11} & 0 \\
m_{21} & m_{22}
\end{pmatrix}
\begin{pmatrix}
b_R^c \\ \tilde{\omega}
\end{pmatrix}+\text{h.c.}
\label{eqn:lbo}
\end{align}
with
\begin{align}
m_{22}=\mu_0-5\mu_1\,,\quad m_{21}=\sqrt{2}\mu_u^3\,, \quad m_{11} = \mu_b\,.
\end{align}
A small bottom mass can be achieved if and
only if
\begin{align}
4 |m_{11} m_{22}| \ll (m_{11}^2+m_{21}^2+m_{22}^2)\,,
\label{eqn:appr}
\end{align}
condition that is compatible with having $\mu_b$ smaller than the other mass terms.
Within the approximation in Eq.~\eqref{eqn:appr}, for small $\mu_b$, we obtain
\begin{align}
\frac{m_b}{\mu_b}  & \approx \frac{|\mu_0 - 5 \mu_1|}{\sqrt{(\mu_0 - 5 \mu_1)^2+2 \mu_{\Delta 2}^2}}\,, \\
M_B & \approx \sqrt{(\mu_0 - 5 \mu_1)^2+2 \mu_{\Delta 2}^2}\,.
\end{align}
The suppression of the bottom mass with respect to $\mu_b$ is thus related to the ratio of masses
\begin{align}
\frac{m_b}{\mu_b}\approx\frac{M_\chi}{M_B}\,,
\end{align}
which is again compatible with the requirement of a light $\chi$. Assuming that $m_b \lesssim \mu_b \lesssim m_t$,
i.e. that the top mass is the largest mass generated by partial compositeness, we obtain the following
range for $M_B$:
\begin{align}
M_\chi \lesssim M_B\lesssim\frac{m_t}{m_b}\times M_\chi\lesssim\frac{m_t}{m_b}\times\Lambda_{\rm HC}\,,
\end{align}
which in turn implies
\begin{align}
|\sqrt{2}\mu_{\Delta 2}|\lesssim\frac{m_t}{m_b}\times\Lambda_{\rm HC}\,.
\end{align}
Namely, $|\mu_{\Delta 2}|$ cannot be too large, otherwise it leads to over-suppression of the bottom mass.
It is also interesting to note the presence of a vector-like bottom quark $B$, with charge $1/3$, which is predicted
to be heavier than the hyper-fermion $\chi$: however, it cannot be much heavier, thus its mass will stay in the multi-TeV range and
$B$ should be discoverable at future high energy colliders.

Finally let us note that when we evolve the PC4F operators from high scale to low scale, radiative corrections
due to hypercharge interaction do not respect $SU(2)_R$ and thus may also contribute to the top-bottom mass
splitting. However, the effect is expected to be small. A naive estimate of the relative correction gives
\begin{align}
\frac{g_Y^2}{(4\pi)^2}\ln\frac{\Lambda_{EHC}}{v_{EW}}\approx 0.05\,,
\end{align}
where $\Lambda_{EHC}\gtrsim 10^{16}\GeV$ denotes the EHC breaking scale, $v_{EW}\approx 246\GeV$ and $g_Y$ is
the hypercharge coupling constant. So we only expect correction at $\mO(10\%)$, which is far from explaining
the complete top-bottom mass splitting.

\subsection{Lepton Masses} \label{sec:leptmass}

As it can be inferred from Eq.~\eqref{eqn:vmpc4f} and Table~\ref{table:PC4Fscalar}, the $\tau$ lepton mass can be generated via
several gauge and scalar-mediated PC4F operators. The model also naturally contains mechanisms that can
explain why leptons are lighter than quarks. From gauge mediation, we saw that lepton PC4Fs are generated
by a different mediator then the quark ones, with a mass that is naturally larger as it is associated to the breaking
of the PS symmetry. If the dominant effect is due to scalar mediators, the masses of the scalars can be
arranged in order to suppress more the lepton operators. In both cases, we also observed that lepton operators
always involve the hyper-fermion $\eta$: if $M_\eta > M_\chi$, therefore, the leptons will be lighter as their mass
is more suppressed. It is, therefore, relatively easy to explain the lightness of the tau with respect to the top.

For neutrinos, the situation is more critical, as they are many orders of magnitude lighter than the corresponding
charged leptons.
If we only consider the effects of PC4F operators, it is possible to generate a neutrino mass that is different
(and suppressed) relative to the charged lepton mass, however it is hard to generate such a large difference
just using the mediator spectra. One possibility could be to rely on the anomalous dimension of the
operator associated to neutrinos.

To make the situation easier, in analogy with the Pati-Salam model, we introduced a singlet fermion $N$~\cite{Volkas:1995yn}. The Yukawa Lagrangian contains
the terms $-\mu_N NN-\lambda_\Phi \Upsilon\Phi N+\text{h.c.}$, the latter of which generates a mixing between 
$N$ and the right-handed neutrino $\nu_{\tau R}^c$ once the scalar $\Phi$ generates the PS-breaking VEV.
This mixing can be used to implement an inverse see-saw mechanism in the model~\cite{Abada:2014vea}.
To illustrate how this works, we will assume that a large Dirac mass is generated for the neutrinos,
in the form $- \mu_\nu\ \nu_L \nu_R^c + \text{h.c.}$, where $\mu_\nu \approx m_\tau$. The singlet $\rho$
also enters in the game via the mixing in Eq.~\eqref{eqn:ly2}. All in all, the relevant mass matrix reads:
\begin{widetext}
\begin{align}
\mL_\nu=-\frac{1}{2} \begin{pmatrix}
\nu_L & \nu_R^c & N & \rho & \tilde{\rho}
\end{pmatrix}
\begin{pmatrix}
0 & \mu_\nu & 0 & 0 & 0 \\
\mu_\nu & 0 & \mu_\Phi & -\mu_{\Delta 1} & 0 \\
0 & \mu_\Phi & \mu_N & 0 & 0 \\
0 & -\mu_{\Delta 1} & 0 & 0 & \mu_0 \\
0 & 0 & 0 & \mu_0 & 0
\end{pmatrix}
\begin{pmatrix}
\nu_L \\ \nu_R^c \\ N \\ \rho \\ \tilde{\rho}
\end{pmatrix}+\text{h.c.}
\end{align}
\end{widetext}
where
\begin{align}
\mu_\Phi \propto \lambda_\Phi v^\Phi_{\rm PS}\,.
\end{align}
As explained in the previous sections, we expect $\mu_{\Delta 1}$ to be relatively small compared to the
scalar VEV scales (it could even vanish in the vacuum with vanishing $\Delta$ VEV), thus we can work
in the approximation where $\rho$ decouples from the rest.
The upper $3\times 3$ block, therefore,  exhibits the inverse seesaw form discussed in Ref.~\cite{Abada:2014vea},
allowing for a small neutrino mass for $\mu_N \ll \mu_\Phi \approx v^\Phi_{\rm PS}$. Other scenarios 
giving realistic neutrino spectra may also be possible.

\subsection{Operator Classification}

In any composite Higgs model with fermion partial compositeness, the onset of a near-conformal dynamics above the
condensation scale is crucial in order to generate an enhanced coupling of the top quark fields. In the TPS model, the
transition between the conformal and confined phases can be traced back to some of the hyper-fermions acquiring
a mass of the order of $\Lambda_{\rm HC}$. Thus, the global symmetries in the two phases are not the same. 
Identifying the operators that couple to the top fields (and to other SM fermions) is crucial in a twofold way: on the
one hand, to be able to check if a sufficient anomalous dimension is generated in the conformal phase; on the
other hand, to identify the hyper-baryons that mix with the SM fermions at low energy. The latter has important consequences
for the low energy phenomenology of the model~\cite{Panico:2015jxa}, and the eventual collider signatures.

We will approach this analysis in the following way:
\begin{itemize}

\item[-] In the conformal window, we identify the operators in terms of the global symmetry $G_{\rm CFT}$, and match
them to the PC4F operators. This allow us to identify the global symmetry properties of each SM fermion partner. The anomalous
dimensions need to be computed on the lattice.

\item[-] At $\Lambda_{\rm HC}$, some heavy fermions can be integrated out, and the low energy theory can be characterized
in terms of ``light'' degrees of freedom, with a global symmetry $G/H$. The SM fermions can now be embedded into 
representations of $G$, while baryons (i.e., spin-1/2 resonances with a definite mass) are matched to the respective
operators and classified in terms of the unbroken symmetry $H$.

\item[-] The low energy effective theory can thus be constructed in terms of the light degrees of freedom, including
light baryon resonances~\cite{Marzocca:2012zn,Panico:2015jxa}.

\end{itemize}
We recall that some fermions, like leptons, may couple to baryons containing a ``heavy'' fermion, i.e. a hyper-fermion with
a mass larger than $\Lambda_{\rm HC}$. In such cases, techniques like HQET~\cite{Eichten:1989zv,Georgi:1990um}, developed
to study bound states containing one bottom or charm quark in QCD, can be deployed.

\begin{table}[t!]
\begin{centering}
\begin{tabular}{|c|c|c|c|c|}
\hline
Field & quantum numbers  & mass & \multicolumn{2}{c|}{collective names}\\
\hline
$L$ & $(\textbf{4}, \textbf{1}, \textbf{2})_0$ & $M_L \ll \Lambda_{\rm HC}$ & \multirow{3}{*}{$\psi_l^i\; [4]$} & \multirow{7}{*}{$\psi^\alpha\; [12]$}\\
$U_1$ & $(\textbf{4}, \textbf{1}, \textbf{1})_{1/2}$ & $M_{R1} \ll \Lambda_{\rm HC}$ &&\\
$D_1$ & $(\textbf{4}, \textbf{1}, \textbf{1})_{-1/2}$ & $M_{R1} \ll \Lambda_{\rm HC}$ &&\\  \cline{4-4}
$U_2$ & $(\textbf{4}, \textbf{1}, \textbf{1})_{1/2}$ & $M_{R2}$ & \multirow{4}{*}{$\psi_h^j\; [8]$} & \\
$D_2$ & $(\textbf{4}, \textbf{1}, \textbf{1})_{-1/2}$ & $M_{R2}$ &&\\
$\eta$ & $(\textbf{4}, \bar{\textbf{3}}, \textbf{1})_{-1/6}$ & $M_\eta > \Lambda_{\rm HC}$ &&\\
$\tilde{\eta}$ & $(\textbf{4}, \textbf{3}, \textbf{1})_{1/6}$ & $M_\eta > \Lambda_{\rm HC}$ &&\\
\hline
$\chi$ & $(\textbf{5}, \textbf{3}, \textbf{1})_{-1/3}$ & $M_\chi \lesssim \Lambda_{\rm HC}$ && \multirow{2}{*}{$\chi^k\; [6]$} \\
$\tilde{\chi}$ & $(\textbf{5}, \bar{\textbf{3}}, \textbf{1})_{1/3}$ & $M_\chi \lesssim \Lambda_{\rm HC}$ &&\\
\hline
\end{tabular}
\caption{Example of ``light'' hyper-fermions in the minimal model, classified in terms of their $(Sp(4)_{\rm HC},\ SU(3)_c,\ SU(2)_L)_{U(1)_Y}$ quantum numbers. The number of Weyl flavors is indicated in square brackets in the ``collective names'' column.\label{tab:HF}}
\end{centering}
\end{table}

In the following we outline the analysis of operator classification according to their transformation properties
under the global symmetry. We simply focus on partners of the left-handed top-bottom doublet, while the analysis for the
remaining quark and lepton partners can be carried out in a similar manner.
The relevant hyper-fermions, with their quantum numbers and collective notations are listed in Table~\ref{tab:HF}. 
The iso-singlet hyper-fermions are indicated in terms of the mass eigenstates, $U_{1,2} \leftrightarrow \{ U_d^3, D_b\}$  and 
$D_{1,2} \leftrightarrow \{ D_u^3, U_t\}$, of the mass matrix in Eq.~\eqref{eq:fr}. For simplicity, we consider that only 4
hyper-fermions in the fundamental of $Sp(4)_{\rm HC}$ are light, togheter with $\chi$, thus they constitute the ``light''
degrees-of-freedom (the other two iso-singlets may also be light, without changing qualitatively the discussion). 
The others have masses of the order of $\Lambda_{\rm HC}$.

In the regime where the hypercolor theory exhibits its strongly-coupled near-conformal dynamics, all hyper-fermions
listed in Table~\ref{tab:HF} are active degrees of freedom. The global symmetry of the composite sector is then
\begin{equation}
G_{\rm CFT} = SU(12)_\psi \times SU(6)_\chi \times U(1)\,,
\end{equation}
where $N_\psi = 12$ and $N_\chi = 6$ count the Weyl spinors in the two species, and $U(1)$ is the anomaly-free abelian
symmetry, with charges $q_{\psi} =- q_\chi = 1$. The spin-1/2 hyper-baryon operators can be constructed with two spinors
of specie $\psi$ and one $\chi$. As to the contraction of
spinor indices, here we note that hyper-baryon operators can be further grouped into two types: $\langle XYZ\rangle$
and $\langle X\bar{Y}\bar{Z}\rangle$, where $X,Y,Z$ are three generic Weyl fermions of the hypercolor group.~\footnote{
We recall that the bar indicates the charge conjugate (right-handed) spinor.}
It is understood that $\langle XYZ\rangle$ contains two irreducible Lorentz representations $(0,1/2)$ and $(0,1/2)'$, while
$\langle X\bar{Y}\bar{Z}\rangle$ contains only one Lorentz representation $(0,1/2)''$~\cite{BuarqueFranzosi:2019eee}.
Note that we focus here on left-handed operators, while right-handed ones can be constructed by replacing each
spinor with its charge-conjugate.
Hyper-baryon operators with definite transformation
properties under the global symmetry group can be constructed schematically as follows:
\begin{eqnarray}
\mathcal{O}_S &=& \frac{1}{2} \langle (\psi^i_\alpha \psi^j_\beta + \psi^j_\alpha \psi^i_\beta) \chi^k_\beta \rangle = (\textbf{S},\textbf{F})_{1}, \label{eq:Os} \\
\mathcal{O}_A &=&\frac{1}{2} \langle (\psi^i_\alpha \psi^j_\beta - \psi^j_\alpha \psi^i_\beta) \chi^k_\beta \rangle = (\textbf{A},\textbf{F})_{1}, \\
\mathcal{O}_{A'} &=& \langle \psi^i_\beta \psi^j_\beta \chi^k_\alpha \rangle = (\textbf{A},\textbf{F})_{1}, \\
\mathcal{O}_{\bar{A}} &=&\langle \bar{\psi}^i_{\dot{\beta}} \bar{\psi}^j_{\dot{\beta}} \chi^k_\alpha \rangle = (\bar{\textbf{A}},\textbf{F})_{-3}, \\
\mathcal{O}_{Adj} &=& \langle \bar{\psi}^i_{\dot{\beta}} \bar{\chi}^k_{\dot{\beta}}  \psi^j_\alpha \rangle = (\textbf{Adj},\bar{\textbf{F}})_{1}, \\
\mathcal{O}_{0} &=& \langle \bar{\psi}^l_{\dot{\beta}} \bar{\chi}^k_{\dot{\beta}}  \psi^l_\alpha \rangle = (\textbf{1},\bar{\textbf{F}})_{1}; \label{eq:O0}
\end{eqnarray}
where $\alpha, \beta, \dot{\alpha}, \dot{\beta}$ are spinorial indices and repeated $\beta$ are contracted with the usual anti-symmetric tensor, 
while $i,j,l$ represent indices of $SU(12)_\psi$ and $k$ of $SU(6)_\chi$.
The notation $(\textbf{S},\textbf{F})_{1}$ means the operator transforms in the two-index symmetric representation
of $SU(12)_\psi$, fundamental representation of $SU(N)_\chi$ and carries a $U(1)$ charge equal to $2q_\psi+q_\chi = 1$.
The meaning of the remaining quantum number notations is self-explanatory. Note also that $\mathcal{O}_A$ and $\mathcal{O}_{A'}$
are the two irreducible Lorentz representations one can build for this type of hyper-baryon operators, while the symmetric $\mathcal{O}_S$ 
can only be constructed with one.
The anomalous dimensions of these operators must be computed on the lattice: yet, as they only depend on the spin and hypercolor
structures, we can derive some interesting relations. First, $\gamma_A = \gamma_S$ and $\gamma_{Adj} = \gamma_0$. Furthermore,
$\mathcal{O}_A$ and $\mathcal{O}_{A'}$ mix as they belong to the same type and have the same charges under the global symmetry~\cite{BuarqueFranzosi:2019eee}.

To match the PC4F operators to the above conformal hyper-baryons, we need to find the correspondence between all 3-fermion
operators that may couple to the SM fields and the operators build above. Below we give an explicit example for the left-handed
quark iso-doublet, with the other cases being straightforward.
All the possibilities are thus listed below:
\begin{equation}
q_L \Rightarrow Q_R^C \rightarrow \left\{ \begin{array}{l}
\left[Q_R^C\right]_{S/A/A'}^1 =  \langle L D_1 \tilde{\chi} \rangle \subset \mathcal{O}_{S/A/A'}\,, \\
\left[Q_R^C\right]_{S/A/A'}^2 =  \langle L D_2 \tilde{\chi} \rangle \subset \mathcal{O}_{S/A/A'}\,, \\
\left[Q_R^C\right]_{S/A/A'}^3 =  \langle L \tilde{\eta} \chi \rangle \subset \mathcal{O}_{S/A/A'}\,, \\
\left[Q_R^C\right]_{\bar{A}}^1 =  \langle \bar{L} \bar{U}_1 \tilde{\chi} \rangle \subset \mathcal{O}_{\bar{A}}\,, \\
\left[Q_R^C\right]_{\bar{A}}^2 =  \langle \bar{L} \bar{U}_2  \tilde{\chi} \rangle \subset \mathcal{O}_{\bar{A}}\,, \\
\left[Q_R^C\right]_{\bar{A}}^3 =  \langle \bar{L} \bar{\eta} \chi \rangle \subset \mathcal{O}_{\bar{A}}\,, \\
\left[Q_R^C\right]_{Adj}^1 =  \langle  \bar{L} D_1 \bar{\chi}\rangle \subset \mathcal{O}_{Adj}\,, \\
\left[Q_R^C\right]_{Adj}^2 =  \langle  \bar{L} D_2 \bar{\chi}\rangle \subset \mathcal{O}_{Adj}\,, \\
\left[Q_R^C\right]_{Adj}^3 =  \langle  \bar{L} \tilde{\eta} \bar{\tilde{\chi}}\rangle \subset \mathcal{O}_{Adj}\,, \\
\left[Q_R^C\right]_{Adj}^4 =  \langle  L \bar{U}_1 \bar{\chi}\rangle \subset \mathcal{O}_{Adj}\,, \\
\left[Q_R^C\right]_{Adj}^5 =  \langle  L \bar{U}_2 \bar{\chi}\rangle \subset \mathcal{O}_{Adj}\,, \\
\left[Q_R^C\right]_{Adj}^6 =  \langle  L \bar{\eta} \bar{\tilde{\chi}}\rangle \subset \mathcal{O}_{Adj}\,.
\end{array} \right.
\label{eqn:qrcop}
\end{equation}
Note the SM gauge quantum numbers should all match. 
The superscript index labels different components inside the same multiplet of the global symmetries that can 
potentially couple to $q_L$: this shows that hyper-baryon operators in the symmetric or antisymmetric have
3 possible ways, while in the adjoint there are 6. 
As mentioned above, the HC dynamics can only mix the two operators $\mathcal{O}_A$ and $\mathcal{O}_{A'}$,
however it will not generate mixing between the various components inside each operator which couple to
the SM fields. This is due to the fact that they are protected by the global symmetries. On the other hand, 
some mixing may be generated by the SM gauge symmetries: this is the
case, for instance, for operators containing $D_{1,2}$ and $U_{1,2}$, as they have exactly the same quantum numbers.
Others cannot mix: for example, we do not expect a mixing between $\left[Q_R^C\right]_{\bar{A}}^3$ and $\left[Q_R^C\right]_{\bar{A}}^{1,2}$,
as the former contain the QCD-charged $\eta$ while the latter contains QCD-neutral iso-singlets.

Vector-mediated PC4F operators associated with $q_L$ can then be classified as
\begin{align}
&\frac{1}{M^2_V} q_L \left[c_1\  \bar{L} \bar{\eta} \chi +c_2^i\  \bar{L} \bar{U}_i \tilde{\chi} +  c_3^j\ \bar{L} \bar{\chi} D_j + c_4\ \bar{L} \bar{\tilde{\chi}} \tilde{\eta} \right] \nonumber \\
& = \frac{1}{M^2_V} q_L \Bigg[c_1\left[Q_R^C\right]_{\bar{A}}^3+c_2^1\left[Q_R^C\right]_{\bar{A}}^{1}+c_2^2\left[Q_R^C\right]_{\bar{A}}^{2} \nonumber \\
& +c_3^1\left[Q_R^C\right]_{Adj}^{1}+c_3^2\left[Q_R^C\right]_{Adj}^{2}+c_4\left[Q_R^C\right]_{Adj}^{3}\Bigg],
\label{eqn:vop}
\end{align}
where the $c_i$'s are calculable dimension-less coefficients.
Note that $c_2^{1,2}$ and $c_3^{1,2}$ are related to each other by rotation angles from Eq.~\eqref{eq:fr}, as they stem
from operators containing $D_b$ and $U_t$ respectively.
For gauge-mediated PC4F operators, therefore, only $\mathcal{O}_{\bar{A}}$ and $\mathcal{O}_{Adj}$ are relevant.
The anomalous dimensions have been computed perturbatively at one loop order in Ref.~\cite{BuarqueFranzosi:2019eee},
yielding:
\begin{eqnarray}
\gamma_{\bar{A}} &=& - \frac{3 g_{\rm HC}^2}{16 \pi^2} \left( 2 C_2 (R_\psi) - C_2 (R_\chi) \right) = - \frac{1}{2} \frac{3 g_{\rm HC}^2}{16 \pi^2}\,,\nonumber \\
\gamma_{Adj} &=& - \frac{3 g_{\rm HC}^2}{16 \pi^2} \left(C_2 (R_\chi) \right) = - 2 \frac{3 g_{\rm HC}^2}{16 \pi^2}\,.
\end{eqnarray}
While these results have limited validity, they seem to suggest the correct sign and that $|\gamma_{Adj}| > |\gamma_{\bar{A}}|$, 
so that the adjoint would lead to larger enhancement.

Once the theory flows down to energies $\sim \Lambda_{\rm HC}$, the heavy hyper-fermions in Table~\ref{tab:HF} can be integrated
out, and the theory with only light flavors condenses and generates dynamically a mass gap. The global symmetry is thus:
\begin{equation}
\frac{G}{H} = \frac{SU(4)_\psi \times SU(6)_\chi \times U(1)}{Sp(4)_\psi \times SO(6)_\chi}\,,
\end{equation}
where the $U(1)$ charges are $q_\psi' = - 3 q_\chi' = 1$. We can now build operators containing two light flavors in the same
way as in Eqs.~\eqref{eq:Os}--\eqref{eq:O0}, except for the the different $U(1)$ charges:
\begin{equation}
\begin{array}{c}
\mathcal{O}_S^{ll} = ({\bf S}, {\bf F})_{5/3}, \;\; \mathcal{O}_{A/A'}^{ll}  = ({\bf A}, {\bf F})_{5/3}, \;\; \mathcal{O}_{\bar{A}}^{ll} = ({\bf \bar{A}}, {\bf F})_{-7/3}, \\
\mathcal{O}_{Adj}^{ll}= ({\bf Adj}, {\bf \bar{F}})_{1/3}, \;\; \mathcal{O}_{0}^{ll}= ({\bf 1}, {\bf \bar{F}})_{1/3},
\end{array}
\end{equation}
where the quantum numbers in parenthesis correspond to the global symmetry $G$.
Operators containing one heavy flavor are also relevant, and they can be classified as:
\begin{eqnarray}
\mathcal{O}_{FF/FF'}^{lh} &=& \langle \psi^l \psi^h \chi \rangle = ({\bf F}, {\bf F})_{2/3}\,, \\
\mathcal{O}_{F\bar{F}}^{lh} &=& \langle \psi^l \bar{\psi}^h \bar{\chi} \rangle = ({\bf F}, {\bf \bar{F}})_{4/3}\,, \\
\mathcal{O}_{\bar{F}F}^{lh} &=& \langle \bar{\psi}^l \bar{\psi}^h \chi \rangle = ({\bf \bar{F}}, {\bf F})_{-4/3}\,, \\
\mathcal{O}_{\bar{F}\bar{F}}^{lh} &=& \langle \bar{\psi}^l \psi^h \bar{\chi} \rangle = ({\bf \bar{F}}, {\bf \bar{F}})_{-2/3}\,.
\end{eqnarray}
The matching of the possible PC4F couplings from Eq.~\eqref{eqn:qrcop} also changes: focusing for simplicity on the
example of the adjoint components in Eq.~\eqref{eqn:vop}, we see
\begin{equation}
[Q_R^C]_{Adj}^1 \subset \mathcal{O}^{ll}_{Adj}\,, \quad  [Q_R^C]_{Adj}^{2,3} \subset \mathcal{O}^{lh}_{\bar{F}\bar{F}}\,.
\end{equation}
This matching allows to construct spurions that encode the SM spinor $q_L$, and can be used to construct the
low energy effective Lagrangian~\cite{Marzocca:2012zn}. As a final step, the operators above should be matched to the
baryon resonances, which have definite masses. They can be classified in terms of the unbroken symmetry
$H$. For instance, 
\begin{equation}
\mathcal{O}^{ll}_{Adj} \to \mathcal{B}^{jj}_{[A,F]} + \mathcal{B}^{jj}_{[S,F]}\,,
\end{equation}
where the subscript denotes the representation under $H = [Sp(4), SO(6)]$. Note that the same hyperbaryon resonance also 
overlap with the other operators, as they share
the same quantum numbers under the unbroken symmetry $H$, but with different structure functions~\cite{Ayyar:2018glg}:
\begin{equation}
\mathcal{O}^{ll}_S \to \mathcal{B}^{jj}_{[S,F]}\,, \;\; \mathcal{O}^{ll}_{A,A',\bar{A}} \to  \mathcal{B}^{jj}_{[A,F]}\,.
\end{equation}
In this case, the most relevant resonance will be determined by the spectrum. In the case of operators containing
one heavy flavor, they all overlap with the same baryon, namely
\begin{equation}
\mathcal{O}^{lh}_{X} \to \mathcal{B}^{lh}_{[F,F]}\,,
\end{equation}
where hyper-baryon operators containing different heavy flavors, $U_2/D_2$, $\eta/\tilde{\eta}$, should be considered as different
states. Also the corresponding baryon resonance will have a mass larger than that of the $\mathcal{B}^{ll}$ states, and
proportional to the mass of the heavy flavour, $M_{R2}$ or $M_\eta$.

\section{Three family model}
\label{sec:ext}

\begin{table*}[tb]
\begin{centering}
\begin{tabular}{|c|c|c|}
\hline
1st Family & 2nd Family & 3rd Family\\
\hline
\phantom{$\begin{array}{c} L\\  L \end{array}$} $N^1$ \phantom{$\begin{array}{c} L\\  L \end{array}$} &
\phantom{$\begin{array}{c} L\\  L \end{array}$} $N^2$ \phantom{$\begin{array}{c} L\\  L \end{array}$} &
\phantom{$\begin{array}{c} L\\  L \end{array}$} $N^3$ \phantom{$\begin{array}{c} L\\  L \end{array}$} \\
\hline
\phantom{$\begin{array}{c} L\\ L\\ L\\ L \end{array}$}
$\Omega^1 = \left( \left( \begin{array}{c} L_u^1 \\ u_L^1 \\ \nu_L^1 \end{array} \right)\; \left( \begin{array}{c} L_d^1 \\ d_L^1 \\ e_L^1 \end{array} \right)\right)$
\phantom{$\begin{array}{c} L\\ L\\ L\\ L \end{array}$} & \phantom{$\begin{array}{c} L\\ L\\ L\\ L \end{array}$}
$\Omega^2 = \left( \left( \begin{array}{c} L_u^2 \\ u_L^2 \\ \nu_L^2 \end{array} \right)\; \left( \begin{array}{c} L_d^2 \\ d_L^2 \\ e_L^2 \end{array} \right)\right)$
\phantom{$\begin{array}{c} L\\ L\\ L\\ L \end{array}$} & \phantom{$\begin{array}{c} L\\ L\\ L\\ L \end{array}$}
$\Omega^3 = \left( \left( \begin{array}{c} L_u^3 \\ u_L^3 \\ \nu_L^3 \end{array} \right)\; \left( \begin{array}{c} L_d^3 \\ d_L^3 \\ e_L^3 \end{array} \right)\right)$
\phantom{$\begin{array}{c} L\\ L\\ L\\ L \end{array}$} \\
\hline
\phantom{$\begin{array}{c} L\\ L\\ L\\ L \end{array}$}
$\Upsilon^1 = \left( \left( \begin{array}{c} U_d^1 \\ d_R^{1c} \\ e_R^{1c} \end{array} \right)\; \left( \begin{array}{c} D_u^1 \\ u_R^{1c} \\ \nu_R^{1c} \end{array} \right)\right)$
\phantom{$\begin{array}{c} L\\ L\\ L\\ L \end{array}$} & \phantom{$\begin{array}{c} L\\ L\\ L\\ L \end{array}$}
$\Upsilon^2 = \left( \left( \begin{array}{c} U_d^2 \\ d_R^{2c} \\ e_R^{2c} \end{array} \right)\; \left( \begin{array}{c} D_u^2 \\ u_R^{2c} \\ \nu_R^{2c} \end{array} \right)\right)$
\phantom{$\begin{array}{c} L\\ L\\ L\\ L \end{array}$} & \phantom{$\begin{array}{c} L\\ L\\ L\\ L \end{array}$}
$\Upsilon^3 = \left( \left( \begin{array}{c} U_d^3 \\ d_R^{3c} \\ e_R^{3c} \end{array} \right)\; \left( \begin{array}{c} D_u^3 \\ u_R^{3c} \\ \nu_R^{3c} \end{array} \right)\right)$
\phantom{$\begin{array}{c} L\\ L\\ L\\ L \end{array}$} \\
\hline
\multicolumn{3}{|c|}{
\phantom{$\begin{array}{c} L\\ L\\ L\\ L \\ L \\ L\end{array}$}
$\Xi = \left( \left[ \begin{array}{c} U_t \\ \chi \\ \eta \\ \omega \\ \rho \end{array} \right] \; \left[   \begin{array}{c} D_b \\ \tilde{\chi} \\ \tilde{\eta} \\ \tilde{\omega} \\ \tilde{\rho} \end{array} \right] \right)$
\phantom{$\begin{array}{c} L\\ L\\ L\\ L \\ L \\ L\end{array}$} }\\
\hline
\end{tabular}
\caption{Extension of the TPS fermion sector to three families. For $\Omega^a$ and $\Upsilon^a$, the two columns correspond to the $SU(2)_{L/R}$ components while the rows are connected by the $SU(8)_{\rm PS}$ symmetry. For $\Xi$, the two columns correspond to the ${\bf 35}$ and ${\bf \bar{35}}$ components of the multiplet under $SU(7)_{\rm EHC}$.
\label{table:3families}}
\end{centering}
\end{table*}

A realistic composite Higgs model must not only account for EWSB within the dynamics of the pNGBs and generate masses for
the third family SM fermions, but also be able to generate masses of the first and second family SM fermions and
 non-trivial mixing matrices. So far, the issue of flavor physics in composite models has been discussed only
in the context of effective field descriptions, for both quarks~\cite{Agashe:2009di,Matsedonskyi:2014iha,Cacciapaglia:2015dsa,Panico:2016ull} and leptons~\cite{Carmona:2013cq,Carmona:2015ena,Frigerio:2018uwx}, or in extra dimensional
holographic descriptions~\cite{Cacciapaglia:2007fw,Fitzpatrick:2007sa,Csaki:2008zd}. Models with a microscopic description of the composite dynamics~\cite{Ferretti:2013kya,Barnard:2013zea}
do not go beyond the generation of the top mass. In particular, in Ref.~\cite{Cacciapaglia:2015dsa} a model was proposed where
two scales are identified: a light one where the physics relevant for the top quark resides with light top
partners, and a larger scale where masses for the light generations and flavor mixing are generated. This
approach has been pushed forward in Ref.~\cite{Panico:2016ull}, where a multi-scale scenario is discussed where each SM
fermion has a partner at a different mass scale. Our PUPC approach offers the \emph{unique} opportunity
to explore \emph{in detail the origin} of flavor physics and fermion masses in a composite Higgs scenario: while
in previous approaches the couplings relevant for flavor physics were added as effective operators, without any
possible attempt to investigate the physics that sources them, in the PUPC approach they can be clearly associated
to either gauge or scalar couplings. They can, therefore, be considered fundamental by all means. As we will
demonstrate in this section, this has important consequences for the low energy physics. In this section we will, therefore,
describe how to expand the TPS model to give mass to first and second generation.

The first obvious step consists in adding new fermions containing the first and second family SM fermions,
in terms of TPS gauge multiplets. The simplest option is to introduce two more copies of $\Omega$ and $\Upsilon$,
see Eqs~\eqref{eq:Omega} and \eqref{eq:Upsilon}. A priori, there is no need to introduce more copies of the
$\Xi$ field since it does not contain SM fermions. The sterile fermion $N$ is also extended to three families.
In Table~\ref{table:3families} we summarize in detail the fermion multiplets and their components. We want
to remark the introduction of two additional copies of the hyper-fermions $L$, $U_d$ and $D_u$, which come along
 the SM fermions. Thus, the total number of hyper-fermions in the fundamental of $Sp(4)_{\rm HC}$
becomes $N_\psi=20$, which is too much in order to keep the theory inside a near-conformal phase
below the TPS symmetry breaking, as discussed in Sec.~\ref{sec:HCdyn}. This observation already suggests
that the hyper-fermions associated to the first two generations should be heavy, with a mass close to
the TPS symmetry breaking scale.~\footnote{The only way to keep all the hyper-fermions light is to break
the CHC symmetry at low energy, close to $\Lambda_{\rm HC}$, so that the conformal window is generated
by the $SU(4)_{\rm CHC}$ dynamics, C.f. Sec.~\ref{sec:HCdyn}.}

The next step consists in extending the Lagrangian to the three family case: adding family indices to Eq.~\eqref{eqn:lyukawa},
we obtain
\begin{align}
\mL_Y & =-\frac{1}{2}\mu_N^{a} N^a N^a-\frac{1}{2} \mu_\Xi \Xi\Xi-\frac{1}{2} \lambda_\Psi \Xi\Psi\Xi - \left( \lambda_\Phi^{ab} \Upsilon^a\Phi N^b \right. \nonumber \\
& \left. + \lambda_{\Theta L}^a\Omega^a\Theta^*\Omega^a + \lambda_{\Theta R}^a\Upsilon^a\Theta\Upsilon^a
+ \lambda_\Delta^a \Upsilon^a\Delta^*\Xi+\text{h.c.} \right)\,,
\label{eqn:lyukawa3fam}
\end{align}
where, without loss of generality, we have used the $U(3)$ flavor symmetry of the fields $N^a$, $\Omega^a$ and
$\Upsilon^a$ to diagonalize the matrices $\mu_N$ and $\lambda_{\Theta L/R}$.
We can already remark that the only terms that connect different flavors are $\lambda_\Phi^{ab}$, which
characterizes the mixing between right-handed and sterile neutrinos, and $\lambda_\Delta^a$, which
introduces couplings between the right-handed SM fermions and the hyper-fermions contained in $\Xi$.

As discussed in the previous section, masses for the hyper-fermions in $\Omega^a$ and $\Upsilon^a$ are
generated by the Yukawas $\lambda_{\Theta L/R}$ upon $\Theta$ developing its CHC-breaking
VEV. Thus, in order to preserve a wide walking window, we need
\begin{equation}
v^\Theta_{\rm CHC} \gg \Lambda_{\rm HC}\,, \;\; \lambda_{\Theta L/R}^{1,2} \sim \mathcal{O} (1)\,, \;\; \lambda_{\Theta L/R}^3 \ll 1\,.
\label{eq:condTheta}
\end{equation}
The latter comes from the need to keep the hyper-fermions of the third generation light, as discussed in the previous
section. Note that this necessary set-up already allows us to rule out
the scenario of Ref.~\cite{Panico:2016ull} in the TPS framework: as partners of the light generations can only contain the
hyper-fermions $L^{1,2}$, $U_d^{1,2}$ and $D_u^{1,2}$, it is not possible to generate hierarchical
masses for them without spoiling the walking in the near-conformal window (this would lead to an
excessive suppression of the top mass).

In the remainder of this section we will focus on the symmetry breaking pattern involving VEVs for the
scalar multiplets $\Phi$, $\Psi$ and $\Theta$, because it allows to preserve baryon number, as we
will discuss later.

\subsection{Scenarios for EWSB with flavor}

In the previous section, the composite Higgs was associated with the hyper-fermions of the third family and
the ones contained in $\Xi$, which need to remain relatively light. As we have shown, it is also necessary to keep
the hyper-fermions of the light generations very heavy. To discuss light generation masses, we need to first
explore how they can couple to the source of EWSB. We envision three potential scenarios:
\begin{enumerate}

\item \emph{Private Higgs} scenario: it may be possible that each family receives the EWSB from a bound
state of the hyper-fermions of the same generation. This scenario has some similarities with the private
Higgs proposed in Ref.~\cite{Porto:2007ed}. As we will explain below, this case should be discarded.

\item \emph{Flavorful Partial Compositeness}: light generation may be connected to their own partners,
i.e. spin-1/2 resonances from the hyper-barion operators of 1st and 2nd generation. As we mentioned, the need
for a walking window implies that the light generation partners should have a fairly large mass, close to
the CHC breaking scale $v_{\rm CHC}^\Theta$. Unless this scale can be pushed to relatively low values,
this scenario seems unlikely because the masses would be excessively suppressed.

\item \emph{Flavored couplings}: the remaining scenario consist in generating couplings for all SM fermions
to the hyper-fermions of the third generation. The flavor structure is thus embedded in the couplings. As we
will see, this scenario requires an extension of the scalar sector as compared to the minimal model of
Sec.~\ref{sec:tps3}.

\end{enumerate}

To better understand why the scenario 1 should be discarded, we need to closely investigate the global
symmetries of the TPS model extended to three generations.
Firstly, for each family, we may introduce a discrete $\mathbb{Z}_2$ symmetry that we name $\mathbb{Z}_{L,p}$
($p$ being the family index), under which all components of the $\Omega^p$ field are odd, while all other fields are even
(including $\Omega^q$, $q\neq p$).
Secondly, for each family, we may introduce a global $SU(2)_{L,p}$ symmetry, which is the simultaneous $SU(2)_L$ rotation
of all components in $\Omega^p$ (while $\Omega^q$ with $q\neq p$ are untouched).
In the minimal model with a single $\Theta$ field, charaterized by the Yukawa terms in Eq.~\eqref{eqn:lyukawa3fam},
all the $\mathbb{Z}_{L,p}$'s are explicitly preserved by the complete Lagrangian of the TPS model, while the $SU(2)_{L,p}$'s are only
broken due to the $SU(2)_L$ gauging.

The mass terms of the SM fermions in the generation $p$ necessarily break both $\mathbb{Z}_p$ and $SU(2)_{L,p}$, or
in other words the private Higgses $H^p$ are charged under these symmetries. In scenario 1, we implicitly assume that
these symmetries are broken spontaneously, leading to the presence of 3 sets of pNGBs due to the breaking of the
global $SU(2)_{L,p}$ symmetries. While one set constitutes the exact Goldstones of the $W^\pm$ and $Z$ bosons,
the others will acquire a mass via the explicit breaking due to the $SU(2)_L$ gauging, and independent of the mass
of the hyper-fermions. This seems to be in contradiction with the decoupling condition~\cite{Preskill:1981sr,Weinberg:1996kr},
which dictates that heavy particles should be decoupled from IR physics. The existence of a massless Goldstone boson
composed of superheavy constituents certainly contradicts the decoupling condition.
Note also that a theorem by Vafa and Witten~\cite{Vafa:1983tf} states that ``non-chiral'' global symmetries cannot
be spontaneously broken. Strictly speaking, the TPS model is not a vector-like theory, even though an $SU(2)_{L,p}$
invariant mass for $L^p$ can be written, so that this theorem cannot be directly applied. Yet, the argument above
suggests that the EWSB must be associated only with light hyper-fermions, i.e. the third generation ones and the ones
contained in $\Xi$, as we studied in the previous section.

Another possibility is that the EWSB is communicated to the heavy hyper-fermions via explicit breaking, like loops
of the $SU(2)_L$ gauge bosons. However, the breaking would be suppressed by the mass of the heavy hyper-fermions,
$\sim v_{\rm SM}^2/v_{\rm CHC}^\Theta$. Unless the CHC breaking scale is low, this possibility is excluded in the same
way as scenario 2.

\subsection{Second Family masses, and the rank of the mass matrix}

In the orginal work proposing Partial Compositeness \cite{Kaplan:1991dc}, D.B.Kaplan realized that, although at
high energy three families with the most general flavor structure are included, the fermion mass
matrix obtained at low energy may turn out to be of rank $1$, as its entries can be expressed as
\begin{align}
m_{ab}=\kappa_a\tilde{\kappa}_b\,,
\label{eqn:entries}
\end{align}
where $a,b=1,2,3$ are family indices.
Thus, to generate masses for the first and second families he introduced mechanisms other than PC.
In the TPS model we should also check that the rank of the mass matrix is enough to give mass to all
generations. For each SM fermion $f$, the mass matrix can be schematically written as
\begin{equation}
M_f=\begin{pmatrix}
<O_{L1}O_{R1}> & <O_{L1}O_{R2}> & <O_{L1}O_{R3}> \\
<O_{L2}O_{R1}> & <O_{L2}O_{R2}> & <O_{L2}O_{R3}> \\
<O_{L3}O_{R1}> & <O_{L3}O_{R2}> & <O_{L3}O_{R3}>
\end{pmatrix}\,,
\label{eqn:massm}
\end{equation}
where $O_{L/Ra}$, with $a=1,2,3$, are the sum of hyper-baryon operators that couple to the SM fermion fields
$f_{Li},f_{Rj}$, while $<...>$ denotes the Fourier-transformed correlator at zero external momenta.

Eq.~\eqref{eqn:massm}, which connects the fermion mass matrix and the hyper-baryon correlator matrix, requires
some technical explanations. In the one family case, the relation between the generated fermion mass and
the corresponding two-point hyper-baryon correlator can be derived by matching the functional derivatives
of the generating functional obtained in the low-energy effective theory (described in terms of pNGBs and
external elementary fields) and the UV description of the
model~\cite{Golterman:2015zwa}. Here we simply generalize the formula to the three family case. Since the
low-energy effective theory is valid up to $\Lambda_{\rm HC}$, the matching must be done at low-energy
as well. To compute the fermion mass matrix $M_f$, therefore, the operators $O_{Li},O_{Rj}$ that appear
in Eq.~\eqref{eqn:massm} should be viewed as renormalized operators defined at $\sim\Lambda_{\rm HC}$.
The running and mixing effects, together with all effects of original couplings and integrating
out mediators, have been taken into account in the definition of these operators.

We note that one PC4F operator can be mediated by multiple vector and scalar mediators. In the scalar mediator part there can be
complicated mixing which affects the mass eigenvalues and Yukawa couplings of the scalar components.
Nevertheless, as long as we go below the scale of the lightest mediator
mass, all PC four-fermion interactions can be incorporated into local effective PC4F operators,
regardless of the origin and properties of the mediators. Moreover, let us note that, mediator masses
and mixings are certainly family-independent, and one side of the mediator must be connected to two
hyper-fermions which is also described by a family-independent coupling. The family-dependence only
comes in at the other side where a scalar mediator is connected to one SM fermion and one hyper-fermion,
and is only embodied in one proportionality factor at tree-level.

Complication may arise due to the hierarchical hyper-fermion masses. When the theory is evolved from
UV to IR, in principle we should integrate out heavy hyper-fermions when we go below the corresponding mass
thresholds. However, if this is done for all hyper-fermions heavier than $\Lambda_{\rm HC}$, Eq.~\eqref{eqn:massm}
may be invalid since some contributions other than hyper-baryon correlators are ignored. On the other hand,
the form of Eq.~\eqref{eqn:massm} is convenient for the analysis of its rank. Our strategy will be as follows.
We subdivide all hyper-fermions heavier than $\Lambda_{\rm HC}$ into two types. The first type includes those
hyper-fermions that are so heavy that their effect on SM fermion mass generation can be safely ignored.
This is the case for hyper-fermions in the first and second families, which are assumed to have superheavy
masses $\sim v_{\rm CHC}^\Theta$. The second type includes those hyper-fermions that have a mass close to
$\Lambda_{\rm HC}$, like $\eta$ and $\tilde{\eta}$, as their effect on SM fermion mass generation cannot be
ignored. We will simply integrate out hyper-fermions of the first type, but retain hyper-fermions of the second type
when we perform the matching to obtain Eq.~\eqref{eqn:massm}. In this manner the convenience of
Eq.~\eqref{eqn:massm} is retained. Of course if concrete calculations are to be carried out,
we need be extremely careful about how the correlators involving heavy hyper-fermions are computed. However
in the following analysis we are not bothered with such complication since we are only concerned with
the rank of $M_f$.

Now, one of the elementary property of the correlator $<O_{La}O_{Rb}>$ is that it is linear with respect
to the participating operators $O_{La}$ and $O_{Rb}$. This sounds trivial but it turns out to be crucial
for the model building. For example, suppose the participating hyper-baryon operators have the structure
\begin{align}
O_{La}=y_{La}O_L,\quad O_{Rb}=y_{Rb}O_R\,,
\end{align}
where $O_L,O_R$ are fermionic operators, and $y_{L/Ra}$ are arbitrary coefficients,
then we immediately realize the resulting mass matrix will have entries like Eq.~\eqref{eqn:entries},
which means its rank is $1$ and will not be able to give masses to all three families.

What is the situation for the TPS model described so far? Firstly, we note that gauge-mediation can
only be effective for third generation, as it only couples components inside the same multiplet. Scalar
mediation, on the other hand, is sensitive to the details of the Yukawa interactions in Eq.~\eqref{eqn:lyukawa3fam}.
The couplings of the left-handed doublets, contained in $\Omega^p$, are only generated by the Yukawa
$\lambda_{\Theta L}^a$, which is diagonal. This implies that only the third generation SM fermions can
couple to the light hyper-fermions, and furthermore, $\lambda_{\Theta L}^3 \ll 1$. Thus, the left-handed
operators will have the form:
\begin{equation}
O_{La} = \delta_{a3} O_L\,,
\end{equation}
leading to rank-$1$ mass matrix.
To mend this problem, we can extend the minimal model by adding a second $\Theta$ scalar and a second
$\Psi$ scalar. We can further use a rotation symmetry between the two to cast the VEVs on the first, $\Theta_1$
and $\Psi_1$, while the second ones, $\Theta_2$ and $\Psi_2$, have a large mass. The Yukawa Lagrangian
now contains two copies of the couplings, as listed below:
\begin{multline}
\mL_Y \supset  - \left( \lambda_{\Theta L}^{kab} \Omega^a \Theta^\ast_k \Omega^b
+  \lambda_{\Theta R}^{kab} \Upsilon^a \Theta_k \Upsilon^b + \mbox{h.c.} \right) \\
-\frac{1}{2} \lambda_\Psi^k \Xi \Psi_k \Xi\,.
\end{multline}
We can again use $U(3)$ flavor rotations to cast $\lambda_{\Theta L/R}^{1ab}$ into a diagonal form,
so that the mass matrices for the hyper-fermions generated by the $\Theta_1$ VEV are diagonal. Note
that $\lambda_{\Theta L/R}^{1ab}$ entries need to fulfil the condition in Eq.~\eqref{eq:condTheta}, while
all the entries in $\lambda_{\Theta L/R}^{2ab}$ can be sizeable. Similarly, we can have $\lambda_\Psi^1 \ll 1$
and $\lambda_\Psi^2 \sim \mathcal{O}(1)$.
From Table~\ref{table:PC4Fscalar}, we see that PC4F operators for $q_L$ and $l_L$ can be generated by the
scalar components $\varphi_2$ and $\varphi_1$ respectively, via mixing between $\Theta^2$ and $\Psi^2$.
In both cases, one additional operator is generated, in the form
\begin{equation}
O_{La}' = \lambda_{\Theta L}^{2a3} \lambda_\Psi^2 \mathcal{O}_{L,\varphi}\,,
\label{eq:OLprime}
\end{equation}
which, once added to the one from vector mediation, gives rank-$2$ to the mass matrix, thus allowing
for the second generation masses.

We finally remark that for right-handed fermions, there are already at least 3 channels: the gauge mediation
for third generation, the $\lambda_{\Theta R}^{2a3} \lambda_\Psi^2$ combination, and the combination
from the $\lambda_{\Delta}$ Yukawa, which can generate at least 3 independent baryonic operators.
In addition, we recall that from Table~\ref{table:PC4Fscalar} the right-handed fermions appear in more
mediator channels than the left-handed ones. The limitation in the rank of the mass matrix, therefore,
uniquely arises from the left-handed sector.

\begin{table}[t!]
\begin{centering}
\begin{tabular}{|c|c|c|c|c|c|c|}
\hline
\multicolumn{7}{|c|}{3-family TPS model} \\
\hline
Field & Spin & $SU(8)_{\rm PS}$ & $SU(2)_L$ & $SU(2)_R$  & $Q_G$ & \# \\
\hline
$\Phi$ & $0$ & $\textbf{8}$ & $\textbf{1}$ & $\textbf{2}$ & $q$ & $1$\\
\hline
$\Theta$ & $0$ & $\textbf{28}$ & $\textbf{1}$ & $\textbf{1}$ & $2q$ & $2$\\
\hline
$\Delta$ & $0$ & $\textbf{56}$ & $\textbf{1}$ & $\textbf{2}$  & $q$ & $1$\\
\hline
$\Delta_L$ & $0$ & $\textbf{56}$ & $\textbf{2}$ & $\textbf{1}$ & $-q$ & $1$\\
\hline
$\Psi$ & $0$ & $\textbf{63}$ & $\textbf{1}$ & $\textbf{1}$  & $0$ &  $2$\\
\hline
$N$ & $1/2$ & $\textbf{1}$ & $\textbf{1}$ & $\textbf{1}$  & $0$ & $3$\\
\hline
$\Omega$ & $1/2$ & $\textbf{8}$ & $\textbf{2}$ & $\textbf{1}$  & $q$ & $3$\\
\hline
$\Upsilon$ & $1/2$ & $\bar{\textbf{8}}$ & $\textbf{1}$ & $\textbf{2}$  & $-q$ & $3$\\
\hline
$\Xi$ & $1/2$ & $\textbf{70}$ & $\textbf{1}$ & $\textbf{1}$ & $0$ &  $1$\\
\hline
\end{tabular}
\caption{Minimal scalar and (left-handed Weyl) fermion field content in the TPS model
that accounts for three families. The last column indicates the minimal number of fields needed.
\label{table:fc3f}}
\end{centering}
\end{table}

\subsection{First family masses}

So far, the first generation of SM fermions remains massless. Adding further $\Theta$ scalar multiplets
does not help: while one can introduce additional flavor structures, they will only appear in a linear combination
to the low energy lagrangian, once the mediators are integrated out. In other words, the form of the
operator in Eq.~\eqref{eq:OLprime} remains unchanged, with $\lambda_{\Theta L}^{2a3}$ replaced by
a linear combination of Yukawa couplings.

A possible solution to this problem consists in introducing a new scalar field $\Delta_L$,
transforming as a $\tbf{56}$ under $SU(8)_{PS}$, doublet under $SU(2)_L$ and singlet under $SU(2)_R$,
and a new Yukawa coupling:
\begin{equation}
\mL_Y \supset - \lambda_{\Delta L}^a \Omega^a \Delta_L \Xi + \mbox{h.c.}
\end{equation}
As $\Delta_L$ is not allowed to develop a VEV, $\lambda_{\Delta L}^a$ can be sizable and generate a new set
of operators for the left-handed doublets, in the form:
\begin{equation}
O_{La}'' = \lambda_{\Delta L}^{a} \lambda_{\Delta L}^3 \mathcal{O}_{L,\Delta}\,,
\label{eq:OLprime2}
\end{equation}
thus elevating the mass matrix rank to the desired $3$. As the flavor structures in the left and right-handed
sectors are independent, this allows to generate the needed flavor mixing and non trivial CKM and PMNS
mixing matrices. CP violating phases can be traced back either to physical phases in the Yukawas or
in phases developed by the hyper-baryon correlators. In Table~\ref{table:fc3f} we summarize the complete
field content of the 3-generation model.

\begin{figure}[tb]
\includegraphics[width=3.0in]{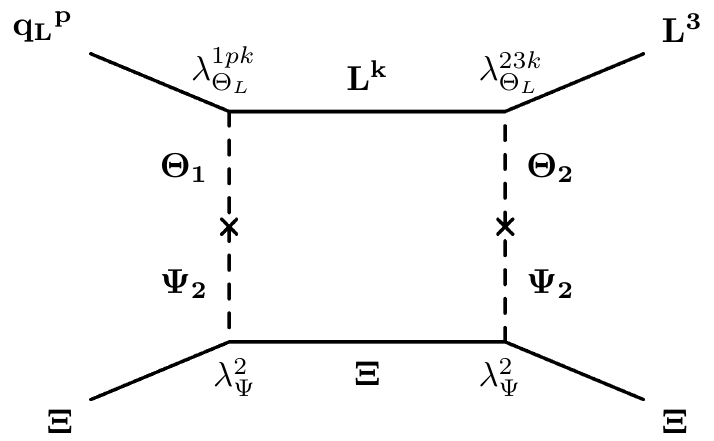}
\caption{\label{fig:looppc4f} Loop-induced PC4F operators as an explanation for the 1st family fermion masses.}
\end{figure}

Another possible solution, which does not require introducing $\Delta_L$, is to consider loop-induced PC4F
operators. This mechanism relies on the fact that the couplings to the superheavy hyper-fermions can
be transmitted to the light hyper-fermions via loops of the Yukawa couplings. As an example, in Fig.~\ref{fig:looppc4f}
we show schematically a loop generating a coupling for the left-handed quarks $q_L$. This would
generate a new coupling of the form:
\begin{equation}
O_{La}'' = (\lambda_\Psi^2)^2 \lambda_{\Theta L}^{1aa} (\lambda_{\Theta L}^{2a3})^\dagger \mathcal{O}_{L,\text{loop}}\,.
\end{equation}
Because of the insertion of $\lambda_{\Theta L}^{1aa}$ (for which $a=1,2$ have large entries, see Eq.~\eqref{eq:condTheta}),
this operator has a different flavour structure than $O_{La}'$ in Eq.~\eqref{eq:OLprime}, thus raising the rank
of the mass matrix to 3, and generating masses for the first generation.

\subsection{Baryon Number conservation and Dark Matter}

In all models where quarks and leptons are unified in a single multiplet, proton decay, or any other
process violating lepton $L$ and baryon $B$ numbers, is a potential threat. Proton and neutron decay experiments,
in fact, can constrain the scale of violation to very high values, $\sim 10^{15\div 16}$~GeV.
The Pati-Salam model~\cite{Pati:1974yy} is known to have neutron-antineutron oscillation instead of proton decay
~\cite{Mohapatra:1980qe,DiLuzio:2011my}. The reason is that, although there exist gauge bosons that
connect quarks and leptons, such transition preserves baryon number. The baryon number violation
then depends on the detail of the scalar sector.

\begin{table}[t!]
\begin{centering}
\begin{tabular}{|l|c|c|c|}
\hline
\multicolumn{4}{|c|}{Global charges} \\
\hline
Fields & $B$ & $L$ & $H$ \\
\hline
SM quarks & $1/3$ & $0$ & $0$ \\
SM leptons & $0$ & $1$ & $0$ \\
$L^p$ & $0$ & $0$ & $1/2$ \\
$U_d^p$, $D_u^p$ & $0$ & $0$ & $-1/2$ \\
\hline
$U_t$ & $-1/2$ & $1/2$ & $1/2$ \\
$\chi$, $\omega$ & $-1/6$ & $1/2$ & $0$ \\
$\eta$ & $1/6$ & $1/2$ & $-1/2$ \\
$\rho$ & $1/2$ & $1/2$ & $-1$ \\
$N^p$ & $0$ & $0$ & $0$ \\
\hline
$v^\Phi_{\rm PS}$ & $0$ & $1$ & $0$ \\
$v^\Psi_{\rm EHC}$ & $0$ & $0$ & $0$ \\
$v^\Theta_{\rm CHC}$ & $0$ & $0$ & $1$ \\
\hline
$v^\Delta_{\rm EHC}$ & $1/2$ & $-1/2$ & $-1$ \\
$v^\Delta_{\rm CHC}$ & $-1/2$ & $1/2$ & $0$ \\
\hline
\end{tabular}
\caption{Global charges $B$, $L$ and $H$ for the fermions in the TPS model. We also list the charges
of the scalar VEVs, to highlight which symmetries are broken. \label{table:U(1)charges}}
\end{centering}
\end{table}

In the TPS model, it is possible to define both ordinary baryon and lepton numbers and a hyperbaryon
number $H$. We normalize $B$ and $L$ like in the SM, while we assign $H$ number $\pm 1/2$ to the
hyper-fermions in the $\Omega^p$ and $\Upsilon^p$ multiplets (see top block in Table~\ref{table:U(1)charges}).
If we only focus on the gauge and Yukawa terms, we realize that $B$, $L$ and $H$ can be consistently assigned
to all the fermion components, as shown in the second block of Table~\ref{table:U(1)charges}.
This can be easily understood by looking at the $U(1)$'s contained in the TPS gauge group: in fact, two
combinations of $B$, $L$ and $H$ are contained in two (broken) generators of $SU(8)_{\rm PS}$ (while
the unbroken hypercharge is defined as a linear combination of $B-L$ inside $SU(8)_{\rm PS}$ and
the diagonal generator of $SU(2)_R$). Finally, the remaining combination corresponds to the global $U(1)_G$
defined in Table~\ref{table:fc3f}, with
\begin{equation}
Q_G = 2H + 3B + L\,,
\end{equation}
which yields $q=1$.
The survival of these symmetries is therefore linked to the breaking of the gauge symmetries: in the bottom
two blocks of Table~\ref{table:U(1)charges}, we report the charges of the VEVs contained in the scalar sector of the theory.
We see that the VEV breaking the $SU(8)_{\rm PS}$ gauge symmetry also violates $L$ (recall that this VEV generates
the mixing between the right-handed neutrinos and the singlets $N$). The CHC breaking VEV in
$\Theta$ breaks $H$ (and generates masses for the hyper-fermions in the $\Omega^p$ and $\Upsilon^p$ multiplets).
Thus, if the breaking is due only to VEVs in $\Phi$, $\Psi$ and $\Theta$, $B$ remains unbroken. Note also that
all the Goldstone bosons associated to the two broken symmetries are eaten by the massive gauge bosons, thus
no light scalar remains. In this section, we will focus on the $B$-preserving scenario, while the $B$-violating case (due
to  the VEVs in $\Delta$) will be discussed in the next subsection. Note finally that no explicit $U(1)_G$ breaking should
be present in the scalar sector.

The main consequence of this scenario, which we shall call \emph{$B$-preserving vacuum}, is that proton and neutron
decays are forbidden, thus avoiding the strong bounds deriving from experiments.~\footnote{Nevertheless, we will
consider that the breaking of the symmetries occurs at high scale, in order to keep the scalar sector ``natural'',
i.e. avoiding a large hierarchy between elementary scalar masses and the Planck scale.} The price to pay is
that mixing between the $\Xi$-components and other fermions are turned off, so that many interesting effects
discussed in Sec.~\ref{sec:tps3}, like the $b_R^c$--$\tilde{\omega}$ mixing, are forbidden. This vacuum, however,
also enjoys the presence of fermions with exotic $B$-charges, which therefore cannot decay back into SM
states. The lightest of the $\Xi$-components, therefore, can play the role of Dark Matter candidate. Of course,
if the lightest state is charged under $Sp(4)_{\rm HC}$, the Dark Matter candidate can be a meson containing
one such hyper-fermion.

\begin{figure}[tb]
\includegraphics[width=2.5in]{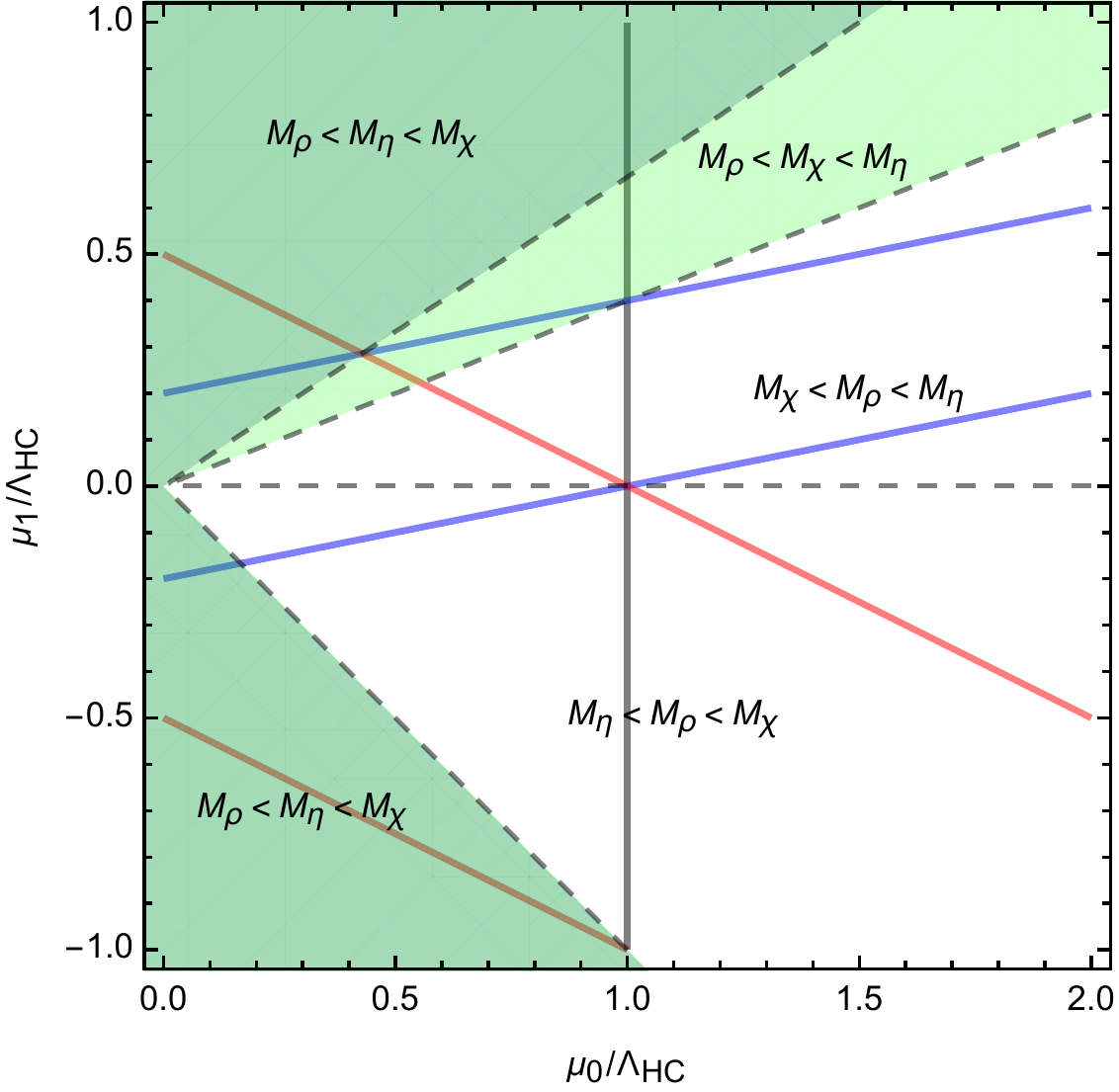}
\caption{\label{fig:masshierarchy} Mass hierarchy between $\rho$, $\chi$/$\omega$ and $\eta$ in the $\mu_0$--$\mu_1$ parameter space.
The green regions are favourable for Dark Matter. The solid lines give, as a reference, the boundaries of the regions where
$M_\rho  < \Lambda_{\rm HC}$ (gray), $M_\chi = M_\omega < \Lambda_{\rm HC}$ (blue) and $M_\eta < \Lambda_{\rm HC}$ (red).}
\end{figure}

The mass spectrum of the $\Xi$ components in the $B$-preserving vacuum has been discussed in Sec.~\ref{sec:hypfmass}:
here we simply recall that
\begin{equation}
M_\omega = M_\chi\ =|\mu_0-5\mu_1|, \;\; M_\rho = \mu_0\;,
\end{equation}
and the two iso-singlet hyper-fermions have the same mass as $\rho$, while $M_\eta=|\mu_0+2\mu_1|$ is correlated with the other two.
As $\omega$ does not carry HC charges, it is crucial that it is not the lightest state.
Furthermore, as $\eta$ is the only hyper-fermion in the fundamental of $Sp(4)_{\rm HC}$ that carries
QCD color charges, mesons containing a single $\eta$ or $\tilde{\eta}$ are not good Dark Matter candidates.
In the following, with the aim of presenting a qualilative discussion of the typical
dark matter phenomenology, let us consider the case in which $\rho,\tilde{\rho}$ act as
the dark matter candidate, with the typical parameter space for masses characterized by
\begin{align}
M_\rho  < M_\omega = M_\chi\,, \text{and} \,\, M_\rho < M_\eta
\end{align}
This configuration occurs in the light and dark green areas in Fig.\ref{fig:masshierarchy} in the $\mu_0$--$\mu_1$ parameter space.
The light green wedge also features $M_\chi < M_\eta$, which could explain the lightness of  leptons with respect to quarks
in the same generation, C.f. Sec.~\ref{sec:leptmass}.
 As a final comment, bound states of $U_t$--$D_b$, if they receive a negative contribution to their mass from the binding energy, may also
be lighter than $\rho$ and play the role of composite Dark Matter candidate.
We also checked that all states with exotic $B$ charges can decay into $\rho$: for instance, $\omega \to \tilde{\rho} + t + \tau^-$,
$\langle L_u^3 \eta \rangle \to \rho + \bar{b}$, and so on. These may be very interesting final states to look for at
the LHC or at future high energy hadron colliders.

A detailed study of the Dark Matter phenomenology of $\rho$ goes beyond the scope of this manuscript, and we leave it
for further exploration. Yet, the most interesting property of this Dark Matter candidate is that it is stable thanks
to the ordinary baryon numbers. Its relic density can, therefore, be linked to that of the ordinary baryons under
some simple assumptions: a) a baryon or lepton asymmetry is generated at scales well above the EWSB scale
(for instance, via leptogenesis~\cite{Davidson:2008bu}); b) the EW phase transition is strong.
Both conditions can be attained in the TPS model: the former via the presence of the heavy sterile neutrinos $N$, the
latter thanks to the presence of additional light pNGBs accompanying the Higgs~\cite{Bian:2019kmg,DeCurtis:2019rxl}.
At the EW phase transition, therefore, the lepton or baryon asymmetries will be re-shuffled between the various
active degrees of freedom in thermal equilibrium. The number of $\Xi$-components in the baryon asymmetry can then be
computed following the procedure delineated in Ref.~\cite{Harvey:1990qw} (see also Refs~\cite{Gudnason:2006yj,Ryttov:2008xe}).
In our case, the $\rho$ and $\omega$ are in thermal equilibrium thanks to the couplings to PC4F operators, which are
enhanced at low energy by the anomalous dimensions, as it can be inferred, for instance, from the gauge-mediation
currents in Eqs~\eqref{eq:JE} and \eqref{eq:JC}. More details on this calculation, and the assumptions adopted, are reported in
the Appendix~\ref{app:DMrelic}. The final result is that the ratio of Dark Matter and baryon densities can be written as
\begin{equation}
\frac{\Omega_{\rm DM}}{\Omega_b} = \frac{M_\rho}{m_N} \left| 2 \sigma_D - 2 \sigma_\eta - 5 \sigma_\chi - \sigma_\omega \right| \,,
\end{equation}
where $\sigma_X$ is a Boltzmann suppression factor which depends on the mass of the particle $X$ and the critical temperature $T_\ast$
of the EW phase transition, defined in Eq.~\eqref{eq:sigmaX}.

\begin{figure}[tb]
\includegraphics[width=2.5in]{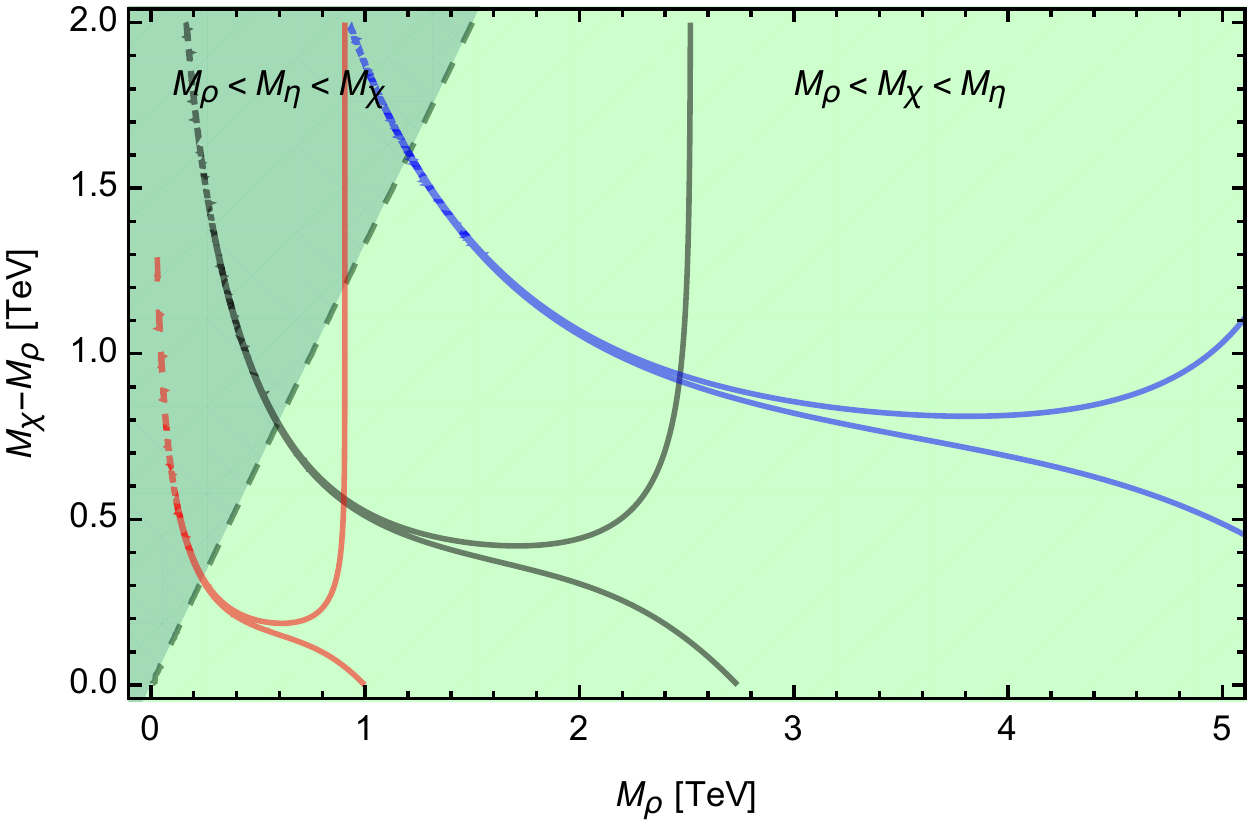}
\caption{\label{fig:DMrelic} Points saturating the DM relic density in the $M_\rho$ vs. $M_\chi - M_\rho$ parameter space. The solid lines correspond to $T_\ast = 246$~GeV
(black), $T_\ast = 100$~GeV (red) and $T_\ast = 500$~GeV (blue).}
\end{figure}

In Fig.~\ref{fig:DMrelic} we show the numerical result in the parameter region where $\rho$ is the lightest $\Xi$-component, focusing on the
$\mu_1>0$ region (C.f. Fig.~\ref{fig:masshierarchy}). We expressed the mass parameters $\mu_{0,1}$ in terms of $M_\rho$ and the mass
difference $M_\chi - M_\rho$, where the dashed line corresponds to $M_\eta = M_\chi$ boundary.
The black line corresponds to points saturating the Planck measurement~\cite{Aghanim:2018eyx} for $T_\ast = v_{\rm SM} = 246$~GeV, showing that the $\rho$
mass is typically between $2.5\div 3$~TeV, except for a funnel region where cancellations between the Boltzmann factors occur. We also show results
for $T_\ast = 100$~GeV (red) and $T_\ast = 500$~GeV (blue), showing how the $\rho$ mass can be lowered or enhanced.
While these results are qualitative, they provide a reliable indication of the typical mass range for the components of the $\Xi$ multiplet, which also have
consequences for the low energy properties of the composite theory. We see that the region with the mass hierarchy $M_\rho < M_\chi < M_\eta$,
relevant in explaining the lightness of lepton masses, seems particularly favourable for this kind of Dark Matter candidate.

\subsection{Baryon Number Violation}

Baryon number violation can occur in the TPS model in two ways: either via explicit interactions
in the scalar potential, or via spontaneous breaking due to scalar VEVs.

As a example of the former, let's consider the following quartic coupling:
\begin{align}
\mL_{V} \supset - \lambda_{4\Theta}\ \epsilon_{ijklmnop}\Theta^{ij*}\Theta^{kl*}\Theta^{mn*}\Theta^{op*}+\text{c.c.}
\label{eq:V4theta}
\end{align}
where $i,\dots ,p$ are $SU(8)_{\rm PS}$ indices. This terms explicitly violates $U(1)_G$, thus it leads to
baryon number violation. If we examine
the decomposition of $\Theta$ at the HC level, we may identify two scalars with quantum numbers
$\theta = (\tbf{1},\tbf{3})_{-1/3}$ and $\bar{\theta} = (\tbf{1},\bar{\tbf{3}})_{1/3}$, which coincides with the quantum number of
one type of scalar leptoquark that can mediate proton decay~\cite{Dorsner:2016wpm}. However, while
$\theta$ has $B_\theta = 1/3$ and $L_\theta = 1$, $\bar{\theta}$ has $B_{\bar{\theta}} = 2/3$ and $L_{\bar{\theta}} = 0$.
Thus, the former behaves like a lepto-quark, while the latter as a di-quark:
\begin{equation}
\theta \to u + e^-\,, \quad \bar{\theta} \to u + d\,.
\end{equation}
The coupling in Eq.~\eqref{eq:V4theta}, after $\Theta$ acquires a VEV, will however generate a mass mixing in the
form $\lambda_{4\Theta} (v^\Theta_{\rm CHC})^2 \theta \bar{\theta}$, thus allowing the standard proton decay operator
\begin{equation}
\frac{1}{M_\theta^2} (u d) (u e^-)\,.
\end{equation}
This kind of processes would require that the mass of these scalars is very large, $M_\theta \approx 10^{15\div 16}$~GeV.

The source of spontaneous $B$-violation is due to VEV(s) for the scalar multiplet $\Delta$, as shown
in the bottom block of Table~\ref{table:U(1)charges}. This scenario has several interesting features,
linked to mixing between the $\Xi$ components and other fermions and hyper-fermions, as discussed
in Sec.~\ref{sec:tps3}. However, it may also generate dangerous $B$ violating effects. One example is
the presence of $B$-violating PC4F operators, mediated by scalars mixing $\Delta$ with other multiplets,
as shown in Table~\ref{table:PC4Fscalar}. Such effects, while suppressed by a large scalar mass, may
be enhanced at low energy by the anomalous running in the conformal phase, thus leaving sizeable
traces at low energy. It may, therefore, not be enough to push the scalar masses and symmetry breaking
scales above the proton decay limits.

\subsection{Final remarks}

We have found that the TPS model can accommodate for masses and flavor mixing between
the three SM generations, once it is suitably extended as shown in Table~\ref{table:fc3f}. The
model can preserve baryon number, $B$, if the symmetry breaking is due to VEVs for $\Phi$,
$\Psi$ and $\Theta^1$, while some couplings in the scalar potential are forbidden. This scenario
also entails a candidate for Dark Matter, protected by a semi-integer baryon number.

One remarkable consequence of the TPS construction is that it fixes many essential properties
of the model in the IR, i.e. in the confined phase. Besides the choice for the HC gauge group,
this goes into the number of light hyper-fermions and their EW quantum numbers. For instance,
we found that the low energy model resembles M8 of~\cite{Belyaev:2016ftv}, except for the hypercharge
of $\chi$ (which is $-1/3$ in the TPS model, instead of $2/3$~\cite{Barnard:2013zea,Ferretti:2013kya}).
This difference implies that the low energy model suffers from corrections to the bottom couplings to
the $Z$ boson~\cite{Agashe:2006at}, with strong bounds on the masses of the baryons as a consequence.

Furthermore, a detailed study of the low energy dynamics is crucial to establish the viability of the model
in view of unwanted flavor and CP violation. This analysis is made more difficult by the ignorance of
the dynamics in the walking phase, which can only be studied on the lattice: although
the flavor scale is superheavy ($\Lambda_{F}\sim 10^{16}\GeV$), flavor-violation is incorporated
into local PC4F operators whose effects are preserved down to $\Lambda_{HC}\sim 10\TeV$ due to large
anomalous dimensions of certain hyperbaryon operators. The flavor-violating couplings are introduced
due to the need to generate masses for the first and second family SM fermions, so we expect flavor
violation is suppressed by light SM fermion Yukawas. However, it is known such suppression is not
enough to be compatible with experimental bounds~\cite{Panico:2015jxa}. CP-violating couplings are
also needed to generate the phase of CKM matrix in order to account for CP-violation phenomena in
the quark sector. However, unwanted CP-violation may result in observables like electron electric
dipole moment (EDM). Recent electron EDM results~\cite{Andreev:2018ayy} lead to strong constraint on
the compositeness scale: $f\gtrsim 100\TeV$ where $f$ is the Goldstone decay
constant~\cite{Panico:2018hal}. In the low-energy effective theory, introducing certain flavor
symmetries may help relax the constraint~\cite{Matsedonskyi:2014iha,Frigerio:2018uwx}. It could be
tricky (if possible) to implement such symmetries in a UV-complete model like TPS, without affecting
generating realistic masses and mixing of SM fermions. We therefore leave this issue for future
study~\cite{Giacomo:2019ehd}.

\section{Summary and Outlook}
\label{sec:dnc}

That EWSB may originate from condensation in a new sector of strong dynamics is an attractive idea.
Compared to the SM Higgs sector, which is parametrized via an elementary scalar field, it may naturally provide
deeper insights into the possible origin of the EWSB and its connection to fermion mass generation. 
With the discovery of a $125\GeV$ Higgs-like particle and the need to accommodate the large top
quark mass, it is then compelling to combine the idea of a pNGB Higgs and fermion partial compositeness
in order to achieve natural and realistic models of EWSB based on strong dynamics. 

In underlying gauge-fermion realisations, PC is realized via four-fermion operators built out
of one SM fermion and three hyper-fermions charged under the new confining HC gauge group. 
In this work, we propose the first complete model, valid up to the Planck scale, that can generate
the necessary four fermion operators (PC4F) in a model that has all the necessary features to 
provide a realistic low energy dynamics. This construction is based on the PUPC framework~\cite{Cacciapaglia:2019dsq},
where the HC and SM gauge symmetries are partially unified. When the larger gauge group undergoes spontaneous symmetry breaking, the resulting
massive gauge bosons (and massive scalars) act as mediators for the PC4F operators. 

Realizing the PUPC framework in practice, however, is highly non-trivial due to the many theoretical and phenomenological
requirements.  We found that the simplest model is
based on an $SU(8)_{\rm PS}\times SU(2)_L\times SU(2)_R$ (TPS) gauge group, which breaks to an $Sp(4)_{\rm HC}$ and
the SM gauge groups at a high scale $\Lambda_{\rm PU}\approx 10^{16}\GeV$. 
A minimal anomaly-free set of fermions can embed both the SM fermions and hyper-fermions needed to generate
PC at low energy. Furthermore, we add suitable scalar fields at high scale (thus being natural) that play the roles
of breaking the gauge group, generate PC4F operators via Yukawa couplings, and give masses to some hyper-fermions.
The last feature is crucial in order to generate a walking dynamics between the UV unified phase and the IR confined
one.
We demonstrated that a renormalizable gauge-Yukawa theory based on the TPS gauge group automatically contains all
the ingredients necessary to achieve the above goals. Thus, by a higher level unification we naturally achieve
a tighter theoretical structure which gives deeper insight of the origin of fermion PC and mass generation.

In this work we have shown how the TPS model can generate masses for the three generations of SM fermions,
with non-trivial mixing among them, while preserving all the attractive features of composite pNGB Higgs models. We
identify several mechanisms that can explain the mass split between the various SM fermions (i.e., leptons versus
quarks, bottom versus top) and the lightness of neutrinos via an inverse see-saw mechanism that arises naturally
in this construction. Finally, the walking phase can be achieved by giving appropriate masses to the hyper-fermions
appearing in the model. 
We pointed out that accidental $U(1)$ symmetries corresponding to the hyperbaryon number, the baryon number and the lepton
number have important and interesting phenomenological consequences. In our TPS construction it is possible to
preserve baryon number, thus avoiding strong constraints from proton and neutron decays, with the bonus feature
of obtaining a Dark Matter candidate thanks to the presence of semi-integer baryon number neutral states.
Under certain circumstances, the relic density can be linked to the baryon asymmetry, leading to typical masses
for the Dark Matter candidate in the few TeV range.

While in this work we have proven the feasibility of the PUPC framework, via the explicit TPS realization, this
work should be considered as a stepping stone to further investigate the phenomenology of the TPS model.
The main points that need further investigations are:
\begin{itemize}
\item We have identified the minimal scalar sector and the phenomenologically relevant symmetry breaking patterns,
due to scalar VEVs that are proven to exist in the literature. It is, nevertheless, necessary to check if the desired VEV
patterns can be realized in the scalar potential of the complete model.

\item The presence of a walking phase, where the theory approaches an IR conformal fixed point, is crucial for the
realization of flavor physics in this model. While estimates seem to support the presence of such a phase in the TPS
model, only lattice calculations can verify this non-perturbatively. Remarkably, in the TPS model both the
gauge symmetry and the fermion properties are specified. Furthermore, calculating the anomalous dimensions of
the hyper-baryon operators in this phase is crucial to understand the flavour structure at low energy.

\item We have shown that the model can generate the needed flavor structures of the SM.  A more detailed analysis
is needed, however, to check if unwanted CP and flavor violating effects survive at low energy, which should
face the strong experimental bounds. This analysis can be done in a reliable way only after lattice input is provided,
in the form of  anomalous dimensions in the walking phase to study the enhancement of flavor violating effects at
low energy, and the spectrum of the baryons below the condensation scale.

\item Finally, the running of the gauge couplings should be studied in detail, in order to check the consistency of 
partial unification, where the QCD and HC ones are the most relevant. This task is daring due to the fact that the HC
dynamics is strong over many decades of energy, thus non-perturbative techniques are needed.

\end{itemize}

Although we do not attempt to solve these issues in the present work, we hope that our model-building effort can
provide new perspectives for understanding and evaluating the pNGB Higgs and PC ideas, and motivate the community
to investigate the related problems and the lattice community to explore uncharted territories that are crucial
for our quest for mass generation.

\subsection*{Acknowledgements}

CZ would like to thank Kingman Cheung, Jean-Pierre Derendinger and Xing-Bo Yuan for helpful discussions.
GC and SV received partial support from the Labex-LIO (Lyon Institute of Origins) under grant ANR-10-LABX-66 (Agence Nationale pour la Recherche), and FRAMA (FR3127, F\'ed\'eration de Recherche ``Andr\'e Marie Amp\`ere'').
GC and SV also acknowledge partial support from the LIA FCPPL (France-China Particle Physics Laboratory) and thank Sun Yat-Sen University, Guangzhou China, for hospitality during the early stages of this work.

\begin{appendix}

\begin{widetext}
\section{Field Decompositions}
\label{app:irreps}

To match the TPS theory in the UV with the composite Higgs model in the IR, it is important 
to  understand the decomposition of the TPS multiplets at various steps of the gauge symmetry 
breaking path. 
To this end, in this appendix we will provide for the reader all the necessary information, following
the steps:
\begin{equation}
SU(8)_{\rm PS} \times SU(2)_R \to SU(7)_{\rm EHC} \times U(1)_E \to SU(4)_{\rm CHC} \times U(1)_Y\,,
\end{equation}
where we omitted the $SU(2)_L$ gauge as it remains unbroken all the way down to the compositeness
scale. Also, we recall that the additional $U(1)_X$ charges, relevant for the $\Psi$--$\Theta$ path, can be
recovered as $Q_X = Q_E - Y$. Also, the $SU(4)_{\rm CHC}$ representations can be easily matched to the
$Sp(4)_{\rm HC}$ ones as follows:
\begin{equation}
{\bf 15}_{\rm CHC} \to {\bf 10}_{\rm HC} \oplus {\bf 5}_{\rm HC}\,, \quad {\bf 6}_{\rm CHC} \to {\bf 5}_{\rm HC} \oplus {\bf 1}_{\rm HC}\,, \quad
{\bf 4/\bar{4}}_{\rm CHC} \to {\bf 4}_{\rm HC}\,.
\end{equation}
To distinguish the components at various steps, we will use the following notation:
\begin{equation}
 \{\textbf{56},\textbf{2}\}\Rightarrow\{SU(8)_{PS},SU(2)_R)\},\quad \tbf{21}_{1/7}\Rightarrow SU(7)_{EHC,U(1)_E}, \quad
[\textbf{1},\bar{\textbf{3}}]_{1/3}\Rightarrow [SU(4)_{CHC},SU(3)_C]_{U(1)_Y}. 
\end{equation}
The decomposition of the $SU(2)_L$ and $SU(2)_R$ gauge bosons being rather straightforward,
we will omit them and report the gauge multiplet of  $SU(8)_{PS}$:
\begin{equation}
\{\tbf{63},\tbf{1}\} = \left\{ \begin{array}{rcl}
\tbf{1}_0 & = & [\tbf{1},\tbf{1}]_0 \\
\tbf{7}_{4/7} & = & [\tbf{1},\tbf{3}]_{2/3} \oplus [\tbf{4},\tbf{1}]_{1/2} \\
\bar{\tbf{7}}_{-4/7} & = &  [\tbf{1},\bar{\tbf{3}}]_{-2/3} \oplus [\bar{\tbf{4}},\tbf{1}]_{-1/2} \\
\tbf{48}_0 & = & [\tbf{1},\tbf{1}]_0 \oplus [\tbf{4},\bar{\tbf{3}}]_{-1/6}\oplus [\bar{\tbf{4}},\tbf{3}]_{1/6}\oplus[\tbf{1},\tbf{8}]_0
 \oplus [\tbf{15},\tbf{1}]_0
\end{array} \right.
\end{equation}

For the scalar fields used in the model building, we have:
\begin{eqnarray}
\Phi = \{\tbf{8},\tbf{2}\} & = & \left\{ \begin{array}{rcl}
\tbf{1}_0 & = & [\tbf{1},\tbf{1}]_0 \\
\tbf{1}_{-1} & = & [\tbf{1},\tbf{1}]_{-1} \\
\tbf{7}_{4/7} & = & [\tbf{1},\tbf{3}]_{2/3} \oplus [\tbf{4},\tbf{1}]_{1/2} \\
\tbf{7}_{-3/7} & = & [\tbf{1},\tbf{3}]_{-1/3} \oplus [\tbf{4},\tbf{1}]_{-1/2} 
\end{array} \right. \\
\Theta = \{\tbf{28},\tbf{1}\} & = & \left\{ \begin{array}{rcl}
\tbf{7}_{-3/7} & = & [\tbf{1},\tbf{3}]_{-1/3} \oplus [\tbf{4},\tbf{1}]_{-1/2}  \\
\tbf{21}_{1/7} & = &[\tbf{1},\bar{\tbf{3}}]_{1/3} \oplus [\tbf{4},\tbf{3}]_{1/6}\oplus [\tbf{6},\tbf{1}]_0 
\end{array} \right. \\
\Delta = \{\tbf{56},\tbf{2}\} & = & \left\{ \begin{array}{rcl}
\tbf{21}_{1/7} & = &[\tbf{1},\bar{\tbf{3}}]_{1/3} \oplus [\tbf{4},\tbf{3}]_{1/6}\oplus [\tbf{6},\tbf{1}]_0  \\
\tbf{21}_{-6/7} & = &[\tbf{1},\bar{\tbf{3}}]_{-2/3} \oplus [\tbf{4},\tbf{3}]_{-5/6}\oplus [\tbf{6},\tbf{1}]_{-1}  \\
\tbf{35}_{5/7} & = & [\tbf{1},\tbf{1}]_1\oplus [\bar{\tbf{4}},\tbf{1}]_{1/2}\oplus [\tbf{4},\bar{\tbf{3}}]_{5/6} \oplus [\tbf{6},\tbf{3}]_{2/3} \\
\tbf{35}_{-2/7} & = &[\tbf{1},\tbf{1}]_0\oplus [\bar{\tbf{4}},\tbf{1}]_{-1/2}\oplus[\tbf{4},\bar{\tbf{3}}]_{-1/6}\oplus [\tbf{6},\tbf{3}]_{-1/3}
\end{array} \right. \label{eq:appDelta} \\
\Delta_L = \{\tbf{56},\tbf{1}\} & = & \left\{ \begin{array}{rcl}
\tbf{21}_{-5/14} & = &[\tbf{1},\bar{\tbf{3}}]_{-1/6} \oplus [\tbf{4},\tbf{3}]_{-1/3}\oplus [\tbf{6},\tbf{1}]_{-1/2}  \\
\tbf{35}_{3/14} & = & [\tbf{1},\tbf{1}]_{1/2}\oplus [\bar{\tbf{4}},\tbf{1}]_{0}\oplus [\tbf{4},\bar{\tbf{3}}]_{1/3} \oplus [\tbf{6},\tbf{3}]_{1/6}
\end{array} \right. \label{eq:appDeltaL}
\end{eqnarray}
while for the adjoint $\Psi = \{\tbf{63},\tbf{1}\}$ the same decomposition as for the $SU(8)_{\rm PS}$ gauge bosons applies.

For the fermion multiplets used in the main text, we obtain
\begin{eqnarray}
\Omega = \{\tbf{8},\tbf{1}\} & = & \left\{ \begin{array}{rcl}
\tbf{1}_{-1/2} & = & [\tbf{1},\tbf{1}]_{-1/2} \\
\tbf{7}_{-3/7} & = & [\tbf{1},\tbf{3}]_{1/6} \oplus [\tbf{4},\tbf{1}]_{0} 
\end{array} \right. \\
\Upsilon = \{\tbf{8},\tbf{2}\} & = & \left\{ \begin{array}{rcl}
\tbf{1}_0 & = & [\tbf{1},\tbf{1}]_0 \\
\tbf{1}_{-1} & = & [\tbf{1},\tbf{1}]_{-1} \\
\tbf{7}_{4/7} & = & [\tbf{1},\tbf{3}]_{2/3} \oplus [\tbf{4},\tbf{1}]_{1/2} \\
\tbf{7}_{-3/7} & = & [\tbf{1},\tbf{3}]_{-1/3} \oplus [\tbf{4},\tbf{1}]_{-1/2} 
\end{array} \right. \\
\Xi = \{\tbf{70},\tbf{1}\} & = & \left\{ \begin{array}{rcl}
\tbf{35}_{-2/7} & = &[\tbf{1},\tbf{1}]_0\oplus [\bar{\tbf{4}},\tbf{1}]_{-1/2}\oplus[\tbf{4},\bar{\tbf{3}}]_{-1/6}\oplus [\tbf{6},\tbf{3}]_{-1/3} \\
\bar{\tbf{35}}_{2/7} & = &[\tbf{1},\tbf{1}]_0\oplus [\tbf{4},\tbf{1}]_{1/2}\oplus[\bar{\tbf{4}},{\tbf{3}}]_{1/6}\oplus [\tbf{6},\bar{\tbf{3}}]_{1/3}
\end{array} \right.  \label{eq:appXi}
\end{eqnarray}

In principle, the multiplet $\Xi$ could be replaced by other anti-symmetric representations of $SU(8)_{\rm PS}$. We will briefly discuss the alternatives below.

\subsection{2-index case}

The fermion multiplet $\Xi$ coule be replaced by a two-index anti-symmetric $\Gamma_2$, and its conjugate $\bar{\Gamma}_2$, decomposing as
\begin{eqnarray}
\Gamma_2 = \{\tbf{28},\tbf{1}\} & = & \left\{ \begin{array}{rcl}
\tbf{7}_{-3/7} & = & [\tbf{1},\tbf{3}]_{-1/3} \oplus [\tbf{4},\tbf{1}]_{-1/2}  \\
\tbf{21}_{1/7} & = &[\tbf{1},\bar{\tbf{3}}]_{1/3} \oplus [\tbf{4},\tbf{3}]_{1/6}\oplus [\tbf{6},\tbf{1}]_0 
\end{array} \right. \\
\bar{\Gamma}_2 = \{\bar{\tbf{28}},\tbf{1}\} & = & \left\{ \begin{array}{rcl}
\bar{\tbf{7}}_{3/7} & = & [\tbf{1},\bar{\tbf{3}}]_{1/3} \oplus [\bar{\tbf{4}},\tbf{1}]_{1/2}  \\
\bar{\tbf{21}}_{-1/7} & = &[\tbf{1},{\tbf{3}}]_{-1/3} \oplus [\bar{\tbf{4}},\bar{\tbf{3}}]_{-1/6}\oplus [\tbf{6},\tbf{1}]_0 
\end{array} \right.
\end{eqnarray}
Comparing with Eq.~\eqref{eq:appXi}, we see that both contain iso-singlet hyper-fermions $D_b$ and $U_t$, QCD-colored hyper-fermions $\eta$--$\tilde{\eta}$, while the new fermions contain two copies of the bottom partners $\omega$--$\tilde{\omega}$. The main difference stands in the $\chi$-sector: for this choice, the $\chi$ has no QCD-colour charges. Thus, all the hyper-baryons coupling to quarks must contain $\eta$ or $\tilde{\eta}$, contrary to what we found in the TPS model with $\Xi$.
Note also that the Yukawa couplings with $\Gamma_2$ would be different from the ones involving $\Xi$.

\subsection{3-index case}

Another alternative consists in using 3-index anti-symmetric representations, which will have the same decomposition as the scalars $\Delta$ and $\Delta_L$. 
In particular, we see from Eq.~\eqref{eq:appDeltaL} that a singlet of the $SU(2)_{L/R}$ would contain a neutral iso-singlet hyper-fermion and a color-triplet with charge $1/6$, which is necessarily stable. To avoid this issue, the minimal option would be to promote the fermion $\Gamma_3$ to a doublet of $SU(2)_R$, thus having the same decomposition as $\Delta$:
\begin{eqnarray}
\Gamma_3 = \{\tbf{56},\tbf{2}\} & = & \left\{ \begin{array}{rcl}
\tbf{21}_{1/7} & = &[\tbf{1},\bar{\tbf{3}}]_{1/3} \oplus [\tbf{4},\tbf{3}]_{1/6}\oplus [\tbf{6},\tbf{1}]_0  \\
\tbf{21}_{-6/7} & = &[\tbf{1},\bar{\tbf{3}}]_{-2/3} \oplus [\tbf{4},\tbf{3}]_{-5/6}\oplus [\tbf{6},\tbf{1}]_{-1}  \\
\tbf{35}_{5/7} & = & [\tbf{1},\tbf{1}]_1\oplus [\bar{\tbf{4}},\tbf{1}]_{1/2}\oplus [\tbf{4},\bar{\tbf{3}}]_{5/6} \oplus [\tbf{6},\tbf{3}]_{2/3} \\
\tbf{35}_{-2/7} & = &[\tbf{1},\tbf{1}]_0\oplus [\bar{\tbf{4}},\tbf{1}]_{-1/2}\oplus[\tbf{4},\bar{\tbf{3}}]_{-1/6}\oplus [\tbf{6},\tbf{3}]_{-1/3}
\end{array} \right.  \\
\bar{\Gamma}_3 = \{\bar{\tbf{56}},\tbf{2}\} & = & \left\{ \begin{array}{rcl}
\bar{\tbf{21}}_{-1/7} & = &[\tbf{1},{\tbf{3}}]_{-1/3} \oplus [\bar{\tbf{4}},\bar{\tbf{3}}]_{-1/6}\oplus [\tbf{6},\tbf{1}]_0  \\
\bar{\tbf{21}}_{6/7} & = &[\tbf{1},{\tbf{3}}]_{2/3} \oplus [\bar{\tbf{4}},\bar{\tbf{3}}]_{5/6}\oplus [\tbf{6},\tbf{1}]_{1}  \\
\bar{\tbf{35}}_{-5/7} & = & [\tbf{1},\tbf{1}]_{-1}\oplus [{\tbf{4}},\tbf{1}]_{-1/2}\oplus [\bar{\tbf{4}},{\tbf{3}}]_{-5/6} \oplus [\tbf{6},\bar{\tbf{3}}]_{-2/3} \\
\bar{\tbf{35}}_{2/7} & = &[\tbf{1},\tbf{1}]_0\oplus [{\tbf{4}},\tbf{1}]_{1/2}\oplus[\bar{\tbf{4}},{\tbf{3}}]_{1/6}\oplus [\tbf{6},\bar{\tbf{3}}]_{1/3}
\end{array} \right. 
\end{eqnarray}
The main drawback of this choice is that is contains a much larger number of hyper-fermions, thus seriously endangering the presence of a walking dynamics in the IR, C.f. sec.~\ref{sec:HCdyn}.

\section{Dark Matter relic density calculation}
\label{app:DMrelic}

\begin{table}[t!]
\begin{centering}
\begin{tabular}{|l|c|c|c|c|c||l|c|c|c|c|c|}
\hline
\multicolumn{6}{|c||}{SM + standard hyper-fermions} &  \multicolumn{6}{|c|}{exotic $B$ fermions} \\
\hline
  &  &  $Q$ & $T_L^3$ & $B$ & $n_f$ &   &  &  $Q$ & $T_L^3$ & $B$ & $n_f$ \\
\hline
$t_L$ & $\mu_{t_L}$ & $2/3$ & $1/2$ & $1/3$ & $9$ & $U_t$ & $-\mu_D$ & $-1/2$ & $0$ & $-1/2$ & $4$ \\
$b_L$ & $\mu_{b_L}$ & $-1/3$ & $-1/2$ & $1/3$ & $9$ & $D_b$ & $\mu_D$ & $1/2$ & $0$ & $1/2$ & $4$ \\
$t_R^c$ & $-\mu_{t_R}$ & $-2/3$ & $0$ & $-1/3$ & $9$ & $\chi$ & $\mu_\chi$ & $-1/3$ & $0$ & $-1/6$ & $15$ \\
$b_R^c$ & $-\mu_{b_R}$ & $1/3$ & $0$ & $-1/3$ & $9$ & $\tilde{\chi}$ & $-\mu_\chi$ & $1/3$ & $0$ & $1/6$ & $15$ \\
$\nu_L$ & $\mu_{\nu_L}$ & $0$ & $1/2$ & $0$ & $3$ & $\eta$ & $\mu_\eta$ & $-1/6$ & $0$ & $1/6$ & $12$ \\
$\tau_L$ & $\mu_{\tau_L}$ & $-1$ & $-1/2$ & $0$ & $3$ & $\tilde{\eta}$ & $-\mu_\eta$ & $1/6$ & $0$ & $-1/6$ & $12$ \\
$\tau_R^c$ & $-\mu_{\tau_R}$ & $1$ & $0$ & $0$ & $3$ & $\omega$ & $\mu_\omega$ & $-1/3$ & $0$ & $-1/6$ & $3$ \\
$\nu_R^c$ & $-\nu_{\nu_R}$ & $0$ & $0$ & $0$ & $3$ & $\tilde{\omega}$ & $-\mu_\omega$ & $1/3$ & $0$ & $1/6$ & $3$ \\
\cline{1-6} 
$L_u^3$ & $\mu_L$ & $1/2$ & $1/2$ & $0$ & $4$ & $\rho$ & $\mu_\rho$ & $0$ & $0$ & $1/2$ & $1$ \\
$L_d^3$ & $-\mu_L$ & $-1/2$ & $-1/2$ & $0$ & $4$ &  $\tilde{\rho}$ & $\mu_\rho$ & $0$ & $0$ & $-1/2$ & $1$ \\
$U_d^3$  & $\mu_U$ & $1/2$ & $0$ & $0$ & $4$ &   &  &  &  &  &   \\
$D_u^3$ & $-\mu_U$ & $-1/2$ & $0$ & $0$ & $4$ &   &  &  &  &   &  \\ 
\hline
\end{tabular}
\caption{Weyl fermions participating to the EW phase transition; $n_f$ indicates the degrees of freedom of each spinor.
\label{table:flist}}
\end{centering}
\end{table}

To compute how the baryon number generated above $\Lambda_{\rm HC}$ is transferred to the SM and to the fermions with fractional baryon number (components of $\Xi$), we consider only the states that have a mass below or around $\Lambda_{\rm HC}$. The fermions are listed in Table~\ref{table:flist}, with their electric charge $Q$, their weak iso-spin $T_L^3$, their baryon number $B$, and the multiplicity (which counts the gauge degrees of freedom). We already imposed the relation between the chemical potentials deriving from the hyper-fermion masses.

We shall also consider the $W^\pm$ gauge boson, for which we choose chemical potential $\mu_W$ associated to $W^-$ (and $-\mu_W$ for $W^+$). The EW interactions within the iso-doublets require:
\begin{equation}
\mu_{b_L} = \mu_{t_L} + \mu_W\,, \quad \mu_{\tau_L} = \mu_{\nu_L} + \mu_W\,, \quad \mu_L = - \frac{1}{2} \mu_W\,.
\end{equation}
To take into account the HC dynamics, which replaces the Higgs sector of the SM, we include in the counting of degrees of freedom the hyper-fermions themselves. This is a rough approximation, as the EW phase transition may occur below the condensation scale, where it would be more appropriate to consider bound states. Nevertheless, as we want to obtain a rough estimate of the Dark Matter mass, to simplify the analysis we will stay within this approximation.

Additional relations between the chemical potentials derive from the PC4F operators that survive at low energy due to the large anomalous dimension enhancement. To simplify the analysis, again, we will only consider gauge-mediated PC4F operators. Looking at the expression of the currents in Eqs~\eqref{eq:JE} and \eqref{eq:JC}, we see that the $\Xi$-components $\rho$ and $\omega$ also participate to PC. Thus, considering the PC4F operators is equivalent to imposing the equality of the chemical potentials of the various components of the currents, namely for $J_E^\mu$:
\begin{equation}
- \mu_{t_L} + \mu_L = \mu_U - \mu_{t_R} = -\mu_U - \mu_{b_R} = -\mu_\chi - \mu_D = - \mu_\eta + \mu_\chi = - \mu_\eta + \mu_\omega = -\mu_\rho + \mu_\eta = -\mu_\omega - \mu_D\,;
\end{equation}
while for $J_C^\mu$:
\begin{equation}
\mu_L + \mu_{\tau_L} = \mu_{\nu_R} - \mu_U = \mu_{\tau_R} + \mu_U = \mu_\eta + \mu_\chi = \mu_\eta + \mu_\omega = \mu_\rho - \mu_D\,.
\end{equation}
The relations above allow to determine all the chemical potentials but 4.

A phase transition of the 1st order is characterized by the vanishing of the total electric charge and iso-spin, given by
\begin{multline}
Q_{\rm tot} = 9 \left[ \frac{2}{3} (\mu_{t_L} + \mu_{t_R}) - \frac{1}{3} (\mu_{b_L} + \mu_{b_R} ) \right] + 3 \left[ - (\mu_{\tau_L} + \mu_{\tau_R}) \right] + 4 \left[ \frac{1}{2} \mu_L 2 \sigma_L + \frac{1}{2} \mu_U 2 \sigma_U + \frac{1}{2} \mu_D 2 \sigma_D \right] + \\
15 \left( - \frac{1}{3} \mu_\chi \right) 2 \sigma_\chi + 12 \left( -\frac{1}{6} \mu_\eta \right) 2 \sigma_\eta + 3 \left( - \frac{1}{3} \mu_\omega \right) 2 \sigma_\omega + 4 (-\mu_W)\,,
\end{multline}
\begin{equation}
T^3_{\rm tot} = \frac{1}{2} \left[ 9 (\mu_{t_L} - \mu_{b_L})  + 3 (\mu_{\nu_L} - \mu_{\tau_R})+ 4 \mu_L 2 \sigma_L \right] - 4 \mu_W\,,
\end{equation}
where we have introduced the statistical factor for fermions
\begin{equation}
\sigma_X = \frac{3}{2 \pi^2} \int_0^\infty dx\; x^2 \cosh^{-2} \left( \frac{1}{2} \sqrt{x^2+z^2} \right)\,, \qquad z = \frac{m_X}{T}\,,
\label{eq:sigmaX}
\end{equation}
$T$ being the temperature.
The conditions $Q_{\rm tot} = 0$ and $T^3_{\rm tot}=0$, together with the EW Sphaleron condition
\begin{equation}
\mu_{t_L} + 2 \mu_{b_L} + \mu_{\nu_L} = 0\,,
\end{equation}
allow to fix all chemical potentials as a function of one.

Finally, the baryon number density in the SM quarks (which corresponds after the EW phase transition to the net baryon number density in the Universe), can be expressed as
\begin{equation}
n_b^{\rm SM} = - \frac{12 (3+\sigma_U)}{6 + 3 \sigma_D + \sigma_\eta + 5 \sigma_\chi + \sigma_\omega} \mu_U\,,
\end{equation}
while the total number density of fermions in the $\xi$--components is
\begin{equation}
n_\Xi = - \frac{12 (3+\sigma_U) (2 \sigma_D - 2 \sigma_\eta - 5 \sigma_\chi - \sigma_\omega)}{6 + 3 \sigma_D + \sigma_\eta + 5 \sigma_\chi + \sigma_\omega} \mu_U = (2 \sigma_D - 2 \sigma_\eta - 5 \sigma_\chi - \sigma_\omega) n_b^{\rm SM}\,.
\end{equation} 
Finally, we can express the relic density of Dark Matter, divided by the baryon density, as
\begin{equation}
\frac{\Omega_{\rm DM}}{\Omega_b} = \frac{M_\rho}{m_N} \left|\frac{n_\Xi}{n_b^{\rm SM}}\right| = | 2 \sigma_D - 2 \sigma_\eta - 5 \sigma_\chi - \sigma_\omega | \frac{M_\rho}{m_N} = 5.36 \,,
\end{equation}
where $m_N \approx 1$~GeV is the nucleon mass, and the numerical value comes from the Planck 2018 measurement~\cite{Aghanim:2018eyx}.
The equation above can be used to determine the mass of the Dark Matter, $M_\rho$, as a function of the temperature of the EW phase transition (which enters in the expressions for the $\sigma$-functions).

\end{widetext}

\end{appendix}


\bibliography{tps_refs}
\bibliographystyle{h-physrev}

\end{document}